%% file: main-TechReport-arXiv.tex
\documentclass[letterpaper,twocolumn,10pt]{article}

\newif\ifSPACEHACK
\SPACEHACKtrue 

\newif\ifDEBUG
\DEBUGfalse

\newif\ifANONYMOUS
\ANONYMOUSfalse

\input{misc/typesetting}

\usepackage{usenix}
\usepackage{multicol}
\usepackage{multirow}

\begin{document}

\input{data/data}

\date{}



\newcommand{\MyTitle}{}

\renewcommand{\MyTitle}{Exploring the Implementation of Software Supply Chain Security Methods in Mitigating Risks: A Case Study of Software Signing}
\renewcommand{\MyTitle}{An Industry Interview Study of Software Signing for Supply Chain Security}

\title{\MyTitle}

\ifANONYMOUS
    \author{Anonymous author(s)}
    
\else
    \author{
    {\rm Kelechi G.\ Kalu}\\
    Purdue University \\ 
    kalu@purdue.edu
    \and
    {\rm Tanmay Singla}\\
    Purdue University \\ 
    singlat@purdue.edu
     \and
    {\rm Chinenye Okafor}\\
    Purdue University \\ 
    okafor1@purdue.edu
     \and
     {} 
     \and
    {\rm Santiago Torres-Arias}\\
    Purdue University \\ 
    santiagotorres@purdue.edu
     \and
    {\rm James C.\ Davis}\\
    Purdue University \\
    davisjam@purdue.edu 
    } 
\fi

\maketitle

\begin{abstract}

\JD{NEWNOTE: The reviewers asked us to change the opening sentence.}
\KC{NEWNOTE: changed the wording as reviewer C asked}
Many software products are composed of components integrated from other teams or external parties.
Each additional link in a software product's supply chain increases the risk of the injection of malicious behavior.
To improve supply chain provenance, many cybersecurity frameworks, standards, and regulations recommend the use of software signing.
However, recent surveys and measurement studies have found that the adoption rate and quality of software signatures are low.
We lack in-depth industry perspectives on the challenges and practices of software signing.

\JD{NEWNOTE: Reviewer did not like us to portray all 18 as high-ranking the way we do here}
\KC{NEWNOTE: fixed}
To understand software signing in practice, we interviewed \interviews experienced security practitioners across \orgs organizations.
We study the challenges that affect the effective implementation of software signing implementation.
We also provide possible impacts of experienced software supply chain failures, security standards, and regulations on \signs adoption.
To summarize our findings:
	(1) We present a refined model of the \sscfm highlighting practitioner's signing practices;
	(2) We highlight the different challenges--technical, organizational, and human--that hamper \signs implementation;
	(3) We report that experts disagree on the importance of signing;
    and
	(4) We describe how internal and external events affect the adoption of \signs.
Our work describes the considerations for adopting software signing as one aspect of the broader goal of improved software supply chain security.

\end{abstract}



%


\section{Introduction}
Like all engineered products, software can be built more quickly when components are reused from other sources~\cite{ossra-2020, kant2023dependencyhell}.
Although this integration accelerates production, it requires trusting external parties~\cite{AJILA20071517}.
This trust increases when the provenance of components is known~\cite{singi_2019_trusted_ssc}, such as via \textit{software signing}. 
Although historically underutilized~\cite{schorlemmer_signing_2024,zahan2023openssf}, 
  due to attacks such as SolarWinds~\cite{growley_navigating_2021},
  signing has become part of many government, academic, and industry proposals to secure software supply chains~\cite{whitehouse_cybersecurity_2021, cooper_protecting_2018, cncf_2021}.

Although software signing has desirable effects, it must be adopted carefully to bring benefits.
At present, measurement studies show that missing or erroneous signatures are common across the software supply chain~\cite{schorlemmer_signing_2024, newman_sigstore_2022, kuppusamy_diplomat_nodate, zahan2023openssf, woodruff_pgp_2023}.
Existing security frameworks recommend software signing, but the general guidelines in these frameworks omit a detailed implementation model~\cite{Pigeon_2022, miranda_zottmann_comparing_2023}.
In order to adopt software signing, organizations would benefit from a detailed understanding of industry practices and challenges. 
This view is currently lacking in the academic and grey literature.

Our work represents the first interview study that describes software signing practices and rationales in industry.
This method gives a deep look at a topic, exposing many directions for further study and implications for practice.
In this work, We interviewed \interviews experienced security practitioners across \orgs organizations.
Our interviews focused on four aspects:
  the software signing practices employed in industry;
  the challenges that hamper the effective implementation of software signing in practice;
  the importance accorded to \signs in practice;
  and
  the influence of failure events and industry standards.
Our results provide a detailed understanding of industry practices and challenges in using software signing to promote software supply chain security.
By integrating our subjects' descriptions, we introduce a refinement of the Software Supply Chain Factory Model that indicates the possible and actual roles of software signing.
Our results reveal three kinds of challenges in adopting signing: technical, organizational, and human aspects.
Additionally, our findings indicate that practitioners hold varying views on the importance of signing for supply chain security.
Lastly, we found that security events --- both major failures such as SolarWinds, and new standards and regulations --- appear to have a minimal effect on the adoption of software signing.

In summary, we make the following contributions:
\noindent
\begin{enumerate}
	\item We refined the software supply chain factory model to reflect diverse industry practices and the forms that signing may take therein (\cref{sec: Rq1_result}).
    \item We identified common technical, human, and organizational challenges in adopting signing (\cref{implem_challenges}).
    \item We report on the relative importance of signing in software supply chain security strategies (\cref{sec:RQ3}).
    \item We discuss the effect that failures, standards, and regulations have on software signing practices (\cref{rq4}).
\end{enumerate}

\vspace{-0.1cm}
\section{\textbf{Background \& Related Works}}
\label{sec:B_and_RW}
\vspace{-0.1cm}

Here, we cover
  software supply chains (\cref{sec:Background-SSC}),
  cybersecurity considerations (\cref{sec:Background-SSCSecurity}),
  and
  relevant empirical studies (\cref{sec:empiricalssc}).

\vspace{-0.1cm}
\subsection{Software Supply Chains} 
\label{sec:Background-SSC}

\begin{figure}
    \centering
    \includegraphics[width=0.90\linewidth]{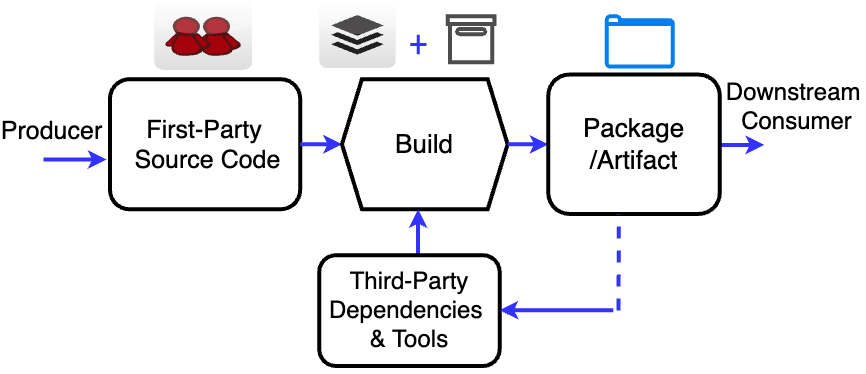}
    \caption{ 
    The Software Supply Chain Factory Model~\cite{SLSA_The_Linux_Foundation_2023}.
    }
    \label{fig:SoftwareFactoryModel}
\end{figure}

Software production is often depicted using variations on the early Software Factory Model~\cite{li_software_2001,cantone_software_1992} or the more recent Software Supply Chain Factory Model (SSCFM)~\cite{SLSA_The_Linux_Foundation_2023, noauthor_tag-securitysupply-chain-securitysecure-software-factorysecure-software-factorymd_nodate}, depicted in~\cref{fig:SoftwareFactoryModel}.
The SSCFM approximates software product from the perspective of one software engineering organization or team
The SSCFM model illustrates that a software product integrates both first-party and third-party components, which are subsequently packaged for downstream use. These integration points also represent potential vulnerabilities where the software product is at risk of malicious code injection from either internal or external sources.

The study of software supply chains has been driven by engineering organizations' greater reliance on external components.
Many works have reported on this reliance in various software contexts.
For example, a 2024 Synopsys report stated that, in 1,067 projects across 17 industry sectors, $\sim$96\% integrated open-source components, and $\sim$77\% of total code is from open source~\cite{OSSRA_2024}.
Even in government and safety-critical systems, the use of third-party open-source components has moved from anathema to normal~\cite{gsa_18f_18f_nodate,department_of_defense_open_2003}.
Analyses of open-source software package registries have shown large increases year-over-year in the number of available components, the number of actively-used components, and the number of transitive dependencies~\cite{wittern2016look,zimmermann_small_nodate, jiang_empirical_2023, nikiforakis_you_2012}.

The increasing reliance on external components in software production amplifies the potential for the exploitation of vulnerabilities that may be contained in these components. As an example of this risk,  Lauinger \etal~\cite{lauinger_thou_2017} report that $\sim$37\% of the most popular Internet websites
rely on at least one library with a publicly acknowledged vulnerability. Given the prevalence of inherited vulnerabilities, Zahan \etal~\cite{zahan_what_2022} and similar studies have summarized attack vectors and potential weak links in software supply chains~\cite{duan_towards_2020, anjum_uncovering_2023, srinivasa_deceptive_2022, yan_estimating_2021, neupane_beyond_2023, gokkaya_software_2023, ohm_towards_2020,ladisa_sok_2023}.

\subsection{Securing Software Supply Chains}
\label{sec:Background-SSCSecurity}


\subsubsection{Methods and Practices} 
\label{sec:methods_practices}
Numerous methods and tools have emerged to improve software supply chain security. 
Okafor \etal~\cite{okafor_sok} state that these approaches promote three distinct dimensions:
 \textit{transparency} (knowing all components in the chain),
 \textit{separation} (isolating components),
 and
 \textit{validity} (ensuring integrity of components).
For example, transparency is increased by adoption of Software Bills of Materials (SBOMs)~\cite{ntia2021sbom,xia_trust_2023}.
Separation is aided by techniques such as
  sandboxing~\cite{sun_nativeguard_2014, abbadini_cage4deno_2023, lamowski_sandcrust_2017, vasilakis_breakapp_2018,vasilakis_supply-chain_2021}
  and
  \JD{NEWNOTE: Update the amusuo to the 2025 ICSE paper reference}
  \KC{NEWNOTE: The reference is not yet online-- i have pinged pascal}
  zero-trust architectures~\cite{stafford2020zero,amusuo2023ztdjava}.
Validity is improved by
  reproducible builds~\cite{lamb_reproducible_2022,vu_lastpymile_2021},
  software signatures~\cite{schorlemmer_signing_2024},
  and
  attestations about artifact metadata (\eg authorship)~\cite{intoto}.
  
Security standards, frameworks, and regulations have been established to complement and direct these security efforts. 
Notable examples include
  The Linux Foundation's Supply Chain Levels for Software Artifacts (SLSA)~\cite{SLSA_The_Linux_Foundation_2023},
  the Cloud Native Computing Foundation's (CNCF) Software Supply Chain Best Practices~\cite{cncf_2021, cncf_supply_chain_security_paper},
  Microsoft's Supply Chain Integrity Model (SCIM)~\cite{SCIM_Microsoft_2023},
  and
  the US National Institute of Standards and Technology's (NIST) Secure Software Development Framework (NIST-SSDF)~\cite{computer_security_division_secure_2021}.
Current national regulations do not force compliance with secure practices~\cite{ludvigsen_preventing_2022}.

\vspace{-0.1cm}
\subsubsection{Software Signing in Detail}
Software signing is a formally guaranteed method of establishing the authorship of a software component~\cite{Garfinkel2003-pa,rfc4880,Internet-Security-Glossary,cooper_security_2018}.
It ensures the provenance and integrity of components in the supply chain through public-key cryptography. The process involves creating a hash of the software, encrypting it with the author’s private key, and attaching the resulting signature to the software. When users download the software, they decrypt the signature with the corresponding public key and compare the hash to ensure the software has not been altered. A typical workflow for a package signing is shown in~\cref{sec:appendix-signing}.

Keyless signing\cite{newman_sigstore_2022, heilman_openpubkey_2023} eliminates the need for authors to manage long-term private keys. Instead, authors authenticate with an identity provider, and an ephemeral key associated with their verified identity is issued for each signing. Users then additionally verify that the public key used for the software verification is associated with the intended author, ensuring both software integrity and authenticity.

Many standards and regulations recommend signing as a base \ssc security technique~\cite{SLSA_The_Linux_Foundation_2023, whitehouse_cybersecurity_2021, souppaya_secure_2022, SCIM_Microsoft_2023,cncf_2021}.  
For example, in The Linux Foundation's SLSA framework, software signing is part of Build Security Level 3~\cite{SLSA_The_Linux_Foundation_2023}.


\vspace{-0.1cm}
\subsection{Empirical SW Supply Chain Security}
\label{sec:empiricalssc}

Although many theoretical techniques for software supply chain security are known, 
adopting them in practice remains a challenge.
Empirical/\textit{in vivo} studies on software supply chain security have examined adoption practices and challenges to identify gaps in theoretical guarantees or usability\cite{whitten1999johnny,ruoti2016johnny, gokkaya_software_2023}. 
Some studies examine the general activities of component management, \eg
  dependency selection~\cite{pashchenko_qualitative_2020} and maintenance~\cite{bogart_when_2021},
  and
  examinations of security challenges and incident handling procedures~\cite{wermke_committed_2022}.
  \JD{NEWNOTE: I feel the next sentence is a bit weak without some citations, since the next *two* sentences imply we just have [39,90] for this claim. Do we have pointers for validity and separation or whatever the other 2 are?}
Others have examined techniques for supply chain security.
For example, for the transparency property ~\cref{sec:methods_practices}, researchers have studied the practices of SBOM adoption~\cite{xia_empirical_2023, kloeg_charting_2024} and reported gaps in tooling, interoperability, adoption, and maintenance. For separation, studies have examined the performance of sandboxing~\cite{wan2019practical} and its adoption in open-source ecosystems~\cite{Alhindi-2024}. For validity, reproducible builds have been analyzed in PyPi~\cite{vu_lastpymile_2021}.

Concerning software signing, researchers have examined challenges in its usability and adoption that create opportunities for attacks~\cite{whitten1999johnny,ruoti2016johnny}.
In 2010–2021, roughly 25\% of publicly reported attacks exploited the missing or erroneous use of signing ~\cite{council2020breaking}.
Top signing-related attack vectors
include self-signed code, broken signature systems, and stolen
certificates ~\cite{gokkaya_software_2023}.
Although software signing is valued, it is considered costly to set up, as reported in Ladisa \etals
survey ~\cite{ladisa_sok_2023,wheeler_2023_SLSA_survey}. Several reports show that signing adoption
rates for open-source components are low. For instance, Zahan \etal found that $\leq$0.5\%  of packages in the npm and PyPI registries had signed releases ~\cite{zahan2023openssf}. Schorlemmer \etal corroborated this proportion, further noting that signing adoption is correlated with registry policies and that the correctness or quality of signatures depends on usable tooling ~\cite{schorlemmer_signing_2024}.
\JD{NEWNOTE: The preceding paragraph is missing any statement about commercial software. If we don't know, write ``we don't know'' and indicate that our paper is the first to shed light. If we do know, say something. But add a statement about the knowledge gap we are filling. We reiterate this in \$3 but we should have some note here as well.}

\vspace{-0.1cm}
\section{\textbf{Knowledge Gaps \& Research Questions}}  ~\label{sec:Research_Questions}
\vspace{-0.1cm}
\label{sec:RQs}



In our analysis of the literature discussed in~\cref{sec:empiricalssc}, we observed that our empirical understanding of software signing challenges and adoption is primarily derived from open-source software components.
The available data on industry practices are surveys~\cite{ladisa_sok_2023,wheeler_2023_SLSA_survey}.
We lack in-depth qualitative descriptions of industry experiences, creating a gap in our understanding of software supply chain security. 
We do not know which signing strategies are effective, the challenges associated with implementing these methods, the roles played by suggested methods and tools, and the factors influencing their adoption.

To fill this gap, we ask four research questions:
\begin{itemize}[leftmargin=33pt, rightmargin=5pt,]
  \setlength{\itemsep}{0pt}  
  \setlength{\parskip}{0pt} 
	\item [\textbf{RQ1:}] Where and how is software signing implemented by software teams? 
	\item [\textbf{RQ2:}] What are the challenges that affect the implementation and use of software signing? 
	\item [\textbf{RQ3:}] What is the perceived importance of software signing in mitigating risks? 
 \item [\textbf{RQ4:}] How do internal and external signing events influence the adoption of software signing? 
\end{itemize}

\section{Methodology} ~\label{sec:methodology}
\begin{figure}
    \includegraphics[width=0.98\columnwidth]{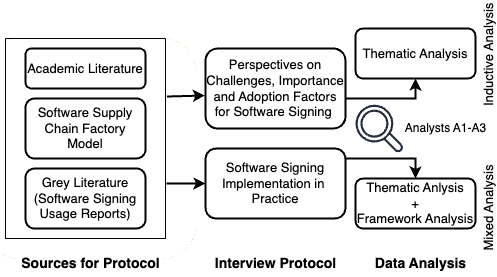}
    \caption{
    Study methodology.
    Several sources (\textit{left}) informed our protocol topics (\textit{center}).
    We applied thematic and framework analyses (\textit{right}).
    \JD{NEWNOTE: there are typos in this figure, eg challlenges and inconsistent cpaitalization. Also add section numbers now that they are stable. And label the magnifying glass with ``3 analysts''?}
    \KC{fixed}
    }
    \label{fig: method}
\end{figure}

\JD{NEWNOTE: In here, I expected us to respond to the concern about using the first few subjects' data, and possibly other concerns from reviewers?}
\KC{fixed}
We conducted semi-structured interviews with subjects who were cybersecurity practitioners.
We depict our methodology in~\cref{fig: method}.
We describe our
  research design (\cref{sec:Approach}), 
  protocol development (\cref{sec:Research_Design}),
  participant recruitment (\cref{sec:Data_Collection}),
  data analysis (\cref{sec:data_analysis}),
  and
  limitations (\cref{sec:Methods-Limitations}).
Our Institutional Review Board (IRB) approved this study.
\JD{NEWNOTE: After we are accepted, add the IRB protocol number. But it is deanonymizing, so not yet.}
\KC{okay}

\subsection{Research Design and Rationale} \label{sec:Approach}

Our research questions require insight into practitioner perspectives.
We considered two factors in selecting a methodology.
First, because the prior work in this space is limited, there was relatively little basis on which to structure a closed-ended survey.
Second, our research questions warrant long-form descriptive-type answers that have to be extracted through open-ended questions, opinions, and detailed observations. 
We therefore opted for a \textit{semi-structured interview}~\cite{saldana2011fundamentals}, which would permit us to capture emergent results~\cite{DeJonckheere2019-ow}.  

The resulting data is expected to be exploratory, and rooted in practitioners' perspectives and experiences~\cite{saldana2011fundamentals}.
The phenomenon of interest is software signing as a security method, which is a contextual decision based on an organization's distinct policies, security expectations, perceived threats, and customer requirements~\cite{sommerville_formal_nodate}.
We therefore adopt an interpretivist perspective for our research methodology design and analysis.
This approach treats each situation as unique~\cite{baltes_sampling_2022},
focusing on in-depth variables and contextual factors~\cite{alharahsheh2020review}. 
Our choice of the interpretivist framework was also motivated by our sampling technique as we discuss later (\cref{sec:Data_Collection}).

\subsection{Protocol Design and Development} ~\label{sec:Research_Design}

We studied three kinds of resources as we created the initial interview protocol.
\begin{enumerate}[leftmargin=*, labelwidth=1.5em, itemindent=0.7em, itemsep=-0.3em]
    \item \textit{Software Supply Chain Security Frameworks:} To ensure a common language and a comprehensive set of topics, we studied software supply chain security frameworks such as SLSA, CNCF, SCIM, and NIST-SSDF (\cref{sec:Background-SSCSecurity}). These frameworks use variants of the Software Supply Chain Factory Model (SSCFM, cf.~\cref{fig:SoftwareFactoryModel}). We therefore used the SSCFM as our reference framework for structuring the C Sections of our interview protocol. This choice allowed us to investigate signing practices at various stages of the SDLC while using a model familiar to our subjects.
    \item \textit{Grey Literature:} To gain insights into practical uses of software signing, we examined grey literature such as industry blogs~\cite{google_software_supply_chain_practices, sigstore_dev}, whitepapers~\cite{cisa_securing_software_supply_chain}, and case studies~\cite{ docker_signing_images, openssf_artifact_attestations, woodruff_pgp_2023}. These sources documented the adoption of software signing in various industry contexts, and, from these, we derived questions about the contextual and organizational factors that subjects consider. We collected these grey literature sources through Web searches of software signing tools mentioned or recommended by any of the security frameworks we studied. Additionally, we reviewed linked articles related to software signing in these standards.
    \item \textit{Background \& Literature Review:} 
    Our background and review of related works (\cref{sec:B_and_RW}) provided a foundation for knowing what questions would be effective for answering our research questions while avoiding duplication. 
    Broadly, our literature review process, which yielded all works cited in~\cref{sec:B_and_RW}) was based on a forward and backward snowball search. We began with works published in prominent cybersecurity and software engineering venues (USENIX, CCS, IEEE S\&P; ICSE, FSE, ASE) and covered a 10-year period.
\end{enumerate}

After protocol development, we refined the protocol through practice interviews (2) and pilot interviews (2).
For practice, the lead author interviewed two of the secondary authors.
These interviews led to modifications in sections B and C of our interview protocol (see~\cref{sec:appendix-Codebooks}).
After recruiting subjects, we treated the first two subjects as pilot interviews.
During these pilots, we gathered additional insights into how subjects perceived software supply chain attacks and reorganized the protocol, particularly focusing on questions related to the perceived sources of greatest risk (section B).
Since few questions were changed or added, we included the data from the pilot interviews in our study where feasible \footnote{We included data from the pilot interviews in our analysis, as qualitative methods guidance supports this practice when the protocol demonstrates its ability to effectively answer research questions ~\cite{Thabane_Ma_Chu_Cheng_Ismaila_Rios_Robson_Thabane_Giangregorio_Goldsmith_2010, holloway1997basic}. This approach is particularly justifiable in contexts where recruitment is challenging, as was the case here and in similar work ~\cite{maxam2024interview}. Quantitatively, we observed that: (1) the pilot interviews used nearly all (~92\%) of the final protocol, as only 3 out of 35 questions were changed afterward, and (2) the unique codes derived from the pilot interview transcripts were comparable in quantity and richness to those from other participants (see ~\cref{fig: saturation}).}.
The final change was made after the seventh interview. The author team reviewed the results up to that point and identified an emergent topic: how practitioners’ organizations integrate feedback into their software supply chain (SSC) security processes. Consequently, one additional question was included in the remaining interviews, to address this emerging topic.

\JD{NEWNOTE: Next paragraph is fine but needs polishing. Possibly better as a footnote on the original text in the original location.}
\KC{fixed}

\noindent
\myparagraph{Defining Internal \& External Events to Answer RQ4}
Unlike the other research questions, RQ4 focuses on investigating how internal and external factors influence organizational decisions to adopt software signing.
Following the definitions of Schorlemmer \etal~\cite{schorlemmer_signing_2024},
we scope this part of the protocol to the impact of (1) \ssc failures, and (2) \ssc standards.

 
\noindent
\cref{tab:InterviewProtocolSummary} summarizes our interview protocol.
The full protocol is given in~\cref{sec:appendix-InterviewProtocol}.

{
\renewcommand{\arraystretch}{1.3}
\begin{table}
    \centering
    \caption{
    Summary of interview protocol.
    Topic D is not analyzed in this paper. See~\cref{sec:appendix-InterviewProtocol} for full protocol.
    \JD{NEWNOTE: It would be better to place the text here into a footnote --- or, ooh, let's put it into Appendix and cref it from here.}
    \KC{fixed}
   }
   \small
   \scriptsize
    \begin{tabular}{
    p{0.32\linewidth}p{0.56\linewidth}
    }
    \toprule
        \textbf{Topic (\# Questions)}& \textbf{Sample Questions} \\
        \toprule
         A. Demographics (4)& {
             What is your role in your team? 
         } \\
         B. Perceived software supply chain risks (4)& {
             Can you describe any specific software supply chain attacks (\eg incidents with 3$^{rd}$-party dependencies, code contributors, OSS) your team has encountered?
         } \\
         C. Mitigating risks with signing (8)& {
             How does the team use software signing to protect its source code (what parts of the process is signing required)?
         }  \\
         \midrule
         D. Signing tool adoption (5)& {
             What factors did the team consider before adopting [TOOL/METHOD] over others?
         }  \\
        \bottomrule
    \end{tabular}
    \label{tab:InterviewProtocolSummary}
\end{table}
}

\subsection{Data Collection}   \label{sec:Data_Collection}

{
	\begin{table}[!t]
		\centering
        \caption{
        Subject Demographics.
        For anonymity, we used generic job roles, not specific titles.
        ``\textit{Senior management}'' refers to senior managers, directors, and executives;
        ``\textit{Technical leader}'' to senior, lead, partner, and principal engineers;
        and
        ``\textit{Engineer}'' and ``\textit{Manager}'' to more junior staff.
        \JD{NEWNOTE: Move the superscripts to just be ``internal'' or ``security''. As long as it fits in the non-diff version we are OK.}
        We also highlight type of software subject's immediate team produces. \textit{*} refers to cases where subject's team built other types of software in addition to the main stated. \textit{O} refers to cases where subject belonged to multiple teams with other products to the one stated. 
        }
             \scriptsize
		\begin{tabular}{llcl}
            \toprule
           \textbf{ID} & \textbf{Role} & \textbf{Experience}  & \textbf{Software Type}\\
            \toprule
            S1 & Research leader & 5 years  & Internal POC software \\
            S2 & Senior mgmt. & 15 years  & SAAS security tool*\textsuperscript{O}\\
            S3 & Senior mgmt. & 13 years  & SAAS security tool*\textsuperscript{O}\\
            S4 & Technical leader  & 20 years & Open-source tooling  \\
            S5 & Engineer   & 2 years & Internal security tooling \\
            S6  & Technical leader & 27 years & Internal security tooling*\textsuperscript{O} \\
            S7 & Manager  & 6 years  & Security tooling*\textsuperscript{O}  \\
            S8 & Technical leader  & 8 years & Internal security tooling*\\
            S9 & Engineer & 2.5 years  & SAAS security*\\
            S10 & Engineer & 13 years & SAAS security* \\
            S11 & Technical leader & 16 years & Firmware* \\
            S12 & Technical leader & 4 years  & Consultancy \\
            S13 & Senior mgmt.  & 16 years & Internal security tool \\
            S14 & Research leader & 13 years  & POC security Software* \\
            S15 & Senior mgmt.  & 15 years & Internal security tooling \\
            S16 & Technical leader & 26 years & Security tooling \\
            S17 & Senior mgmt.  & 15 years& SAAS \\
            S18 & Manager  & 11 years  & Security tooling \\
            \bottomrule
		\end{tabular}
		\label{tab:subjects}
	\end{table}
}

\myparagraph{Target Population} 
To effectively address our research questions, the target population was industry professionals with expertise in software signing.
An ideal subject would have deep knowledge and experience related to software supply chain practices within their organizations.
Such subjects are typically middle- or high-ranking engineering staff.

\noindent
\myparagraph{Recruitment} 
Given the seniority and expected workload of the target population, we anticipated recruitment to be a challenge.
To mitigate this, we employed a non-probability-based purposive and snowball sampling approach to recruit participants~\cite{baltes_sampling_2022}.
This approach incurs the risk of bias, but has the reward of access.
Given the understudied topic at hand, we felt that the reward outweighed the risk.

Our initial sample pool of subjects was made up of members of the Kubecon 2023 conference organizing committee of the Cloud Native Computing Foundation.
This group matches our target population: all of them either worked as a member of the software security teams of their organizations or worked in organizations with products and services in the domain of software security, and all were medium- to high-ranking members of their organizations.
Many of these individuals also contribute to software supply chain security projects hosted by The Linux Foundation, making them well-suited to provide expert insights.
We had special access to this group because one of the authors was a member of that committee.
However, recognizing the potential biases of this initial sample, we expanded our recruitment through recommendations from initial participants and other industry contacts.
This approach allowed us to diversify our subject pool and gather a broader range of insights, while still focusing on individuals with substantial expertise in the domain.

In total, we contacted 30 candidates, of whom \interviews agreed to participate (60\% response rate).
These subjects came from \orgs different organizations.
The subjects were affiliated with companies with security products or were part of their company's security team.
Each participant was offered a \$100 gift card as an incentive, in recognition of their high rank.

\noindent
\myparagraph{Subject Demographics}
We summarize the demographic and organization information of the subjects in~\cref{tab:subjects} and ~\cref{tab:orgSummary}.
We note that six participants were from the initial KubeCon committee invitations; industry contacts of the authors yielded seven participants; and snowball sampling (recommendations from participants) yielded five participants.

In addition to the highlighted, we also note that 13 subjects (from 11 organizations -- 4 Small, 2 Medium, and 6 Large organizations) confirmed that they either lead, or belong to, teams involved in software supply chain, infrastructure and tooling, or security control implementation. These subjects indicated that they are responsible for initiating or implementing their organization's security controls and have responsibilities related to compliance. 
Although not all subjects directly owned compliance, their involvement in security strategy and operational decision making provided them with the appropriate context to assess the utility, influence behind adoption, and challenges affecting the implementation of \sign.

Subjects from small organizations typically focused on SAAS products related to software supply chain, container, and cloud security; or were consultants. In medium-scale organizations, subjects had more specialized roles, \eg developing internal security-enabling software for infrastructure. In large organizations, subjects had similar goals, but often belonged to/oversaw multiple teams.

For confidentiality, we refer to organizations by the letters A--L and participants using the notation SX(y), where \( X \) denotes a unique subject number and \( y \) either distinguishes participants from the same organization or represents the subject's product type. This notation is also used to provide context to subject's responses. For example, \Subjectten refers to Subject S10's direct team's software product, which is a SaaS-type product for an organization whose product area is developer tools.

\myparagraph{Addressing Potential Bias}
 While the connection to the CNCF and specific signing projects could bias responses towards certain approaches, this potential bias is balanced by the deep involvement of participants in the field of software supply chain security. Their contributions are crucial for understanding the current state and future directions of software signing practices. Moreover, the insights provided by these experts, who are at the forefront of the industry, offer a unique and valuable perspective that is essential for the study's objectives.

\myparagraph{ Generalizability of Results}
We acknowledge that the non-random sampling method and the concentration of participants from a specific initiative may limit the generalizability of our findings to the broader population of industry practitioners. However, the purpose of this study was not to achieve broad generalizability but rather to gain in-depth insights from a highly specialized group of experts—hence the use of an interpretivist analysis framework. The findings are particularly relevant to organizations and stakeholders involved in similar software supply chain security initiatives and may inform best practices and future research in this specific context. While broader generalization may be limited, the detailed, expert-driven analysis provided by this study offers significant value in advancing understanding within this niche but critical area.

\myparagraph{Interviews} 
Interviews were conducted by the lead author via Zoom.
The median interview duration was 50 minutes.

\subsection{Data Analysis}  \label{sec:data_analysis}


Interview recordings were transcribed by \url{www.rev.com} (human transcription).
We anonymized identifying information. 
Our analysis then proceeded via multi-stage coding ~\cite{campbell2013coding} and a subsequent thematic ~\cite{braun_using_2006} and framework analysis ~\cite{srivastava_framework__2009}.

Three analysts (A1–A3) participated in the analysis stage of the transcripts. Analyst A1, the lead author, was responsible for all stages of data collection (including protocol design and interview conduct) and contributed to all phases of the analysis.
Initial analysis was conducted by A1 and A2.
When A2 became unavailable, we recruited A3.
To ensure consistency, A3 was trained by memoing and coding transcripts that had already been analyzed.
Comparing A3's memos to A1 and A2, we observed strong alignment, ensuring reliability in the analysis process.
\JD{NEWNOTE: Fix the next sentence. Describe domain expertise, not methods expertise!!!}
\KC{fixed}
A1-A3 are graduate students with research experience in cybersecurity, including software signing. The supervisors are researchers in software engineering and cybersecurity.


\myparagraph{Multi-stage coding}

\vspace{0.08cm}
\noindent
\textit{Stage 1.}
We began with data familiarization~\cite{terry2017thematic}.

\vspace{0.08cm}
\noindent
\textit{Stage 2.}
Next, we needed to develop and refine our codebook.
To provide reliability while conserving resources,
we used the technique described by Campbell \etal ~\cite{campbell2013coding} and  O'Connor \& Joffe ~\cite{oconnor_intercoder_2020}.
First, we chose a proportion of random transcripts to multiply-code (\ie multiple analysts coding).\footnote{Although there is no common agreement on a suitable proportion of the data set to select for this, O'Connor and Joffe recommend 10–25\% as typical~\cite{oconnor_intercoder_2020}. 
We randomly selected approximately 33\% (6 transcripts) to develop our codebook.
This decision to slightly exceed O'Connor and Joffe's recommended percentage was to familiarize analysts A2 and A3 with the nature of our collected data, proposed memo, and coding formats.}
Then, two analysts (A1, A2) conducted multiple rounds of memoing on these transcripts\JD{KELECHI: I added the preceding ``on these transcripts'' which I assume is correct, please check}, discussing emerging code categorizations after each round. This process resulted in 506 coded memos, which were reviewed by Analysts A1 and A3 in a series of meetings.
Ultimately we obtained a codebook with 36 code categories.
This codebook was further refined and extended as discussed in Stages 3-5.  


\vspace{0.08cm}
\noindent
\textit{Stage 3.}
To assess the reliability of our coding, we combine the techniques of Campbell \etal~\cite{campbell2013coding} and Maxam\&Davis~\cite{maxam2024interview}.
Analysts A1 and A3 randomly selected and coded a transcript independently~\cite{campbell2013coding, oconnor_homesnitch_2019}, and we measured their degree of agreement.
Since our codebook was collectively defined by analysts A1 and A3, we supposed this coding task would not be difficult, and thus we used the percentage agreement metric as recommended by Feng~\cite{feng_intercoder_2014}.
In this independent coding step, we observed an 89\% agreement rate.
The analysts were able to resolve disagreements through discussion and code refinement.
According to Campbell \etal~\cite{campbell2013coding}, this high level of agreement suggests that a coding scheme is suitable for a single coding of the remaining uncoded transcripts.



\vspace{0.08cm}
\noindent
\textit{Stage 4.}
Next, a second set of 6 transcripts was selected and coded by Analyst A1 using the resultant codebook from Stage 3.
Following Campbell \etals recommendation~\cite{campbell2013coding}, we created new code categories as they emerged; this phase added 6 new code categories, each with less than 6 associated coded memos.
Analyst A3 reviewed this updated codebook.
\vspace{0.08cm}
\noindent
\textit{Stage 5.}
The same process was applied to the final 6 transcripts, yielding 4 more code categories with fewer than 4 coded memos each. 
Of the 10 new categories recorded from stage 4 and 5, only 3 provided new insights into our data, as determined through multiple review and discussion rounds by Analysts A1 and A3.
Thus A1 and A3 agreed that these new codes be classified as minor code categories. 

\myparagraph{Thematic and Framework Analyses}
From this process, we induced themes following Braune \& Clark's method~\cite{braun_using_2006}.
These themes helped us answer RQs 2-4 directly.
To answer RQ1 we also applied a framework analysis~\cite{srivastava_framework__2009}, mapping the themes to the software supply chain factory model as a reference framework to analyze participants' responses and further categorize the themes. 
This framework was selected because it serves as the general reference for SSC security. 

\JD{Forward cref to the section or table where they are given?}
Through our coding phase’s thematic analysis, we identified eight themes ~\cref{tab:codebook}. Four of these themes were focused on the practical implementation of software signing, while the remaining four addressed various aspects of software signing: the challenges encountered, its perceived importance, the impact of security failures on its adoption, and the influence of security standards and regulations on its adoption.



To determine whether additional aspects of the phenomenon might be expected to emerge if we continued to sample, we measured saturation in the 18 interview transcripts.
Following Guest \etals~\cite{guest2006many} recommendation, we measured saturation  by tracking the cumulative appearance of new codes in each interview.
Each interview was rich, with a median of 33 unique codes (\cref{fig: saturation}).
As shown in~\cref{fig: saturation}, saturation was achieved after interview 11. 

\subsection{Limitation and Threats to Validity} \label{sec:Methods-Limitations}

We discuss construct, internal, and external threats to validity~\cite{wohlin2012experimentation}.
Following the guidance of Verdecchia \etal~\cite{verdecchia2023threats}, we focus on substantive threats that might influence our findings.
We also describe the role that our own perspectives and experiences may have played in this research (positionality)~\cite{dodgson2019reflexivity}.

\ul{Construct Threats} are potential limitations of how we operationalized concepts.
To mitigate these threats, we grounded our interview protocol in the academic and grey literature on software signing.
Following best practices, we also piloted the interview protocol internally and externally to assess its utility in capturing the desired data~\cite{chenail2011interviewing}.

\ul{Internal Threats} are those that affect cause-effect relationships.
Qualitative data implies subjective analysis.
To improve the reliability of our results, we followed an iterative process to develop the codebook and used multiple raters with a high measured level of agreement.
We also acknowledge the possibility of a response bias, as some participants were recruited from the professional network of one of the authors. To address this, the involved author was not present during the interviews and played no role in the data analysis process.

\ul{External Threats} may impact generalizability.
This work is qualitative, which inherently carries limitations. 
Given the cost of an interview study, our work has a smaller sample size (N=18) than methods such as surveys can achieve.
Our sample size is consistent with that of similar studies~\cite{xia_empirical_2023,kloeg_charting_2024}. 
Our subject recruitment approach may also have introduced biases, \eg towards CNCF/The Linux Foundation-oriented frameworks and tools.
As a trade-off, using CNCF members as a starting point improved our access to high-ranking engineering staff.
To reduce bias, we also recruited through snowball sampling and other professional networks.
We assessed this risk with saturation --- by the 11th interview our data saturated, supporting the adequacy of our sample size (\cref{fig: saturation}).
We also segmented our interview protocol (\cref{tab:InterviewProtocolSummary}) and intentionally excluded the tool-specific analysis from this study.
Instead, we concentrated on parts of the protocol related to the general practice of software signing.
\JD{NEWNOTE: Reply to Reviewer A's concern about using expert vs. average subjects here, briefly. Do it in a separate 2-3 sentence paragraph.}

Our study focuses on an expert population to capture nuanced insights into software signing strategies and decisions. While this approach excludes the perspectives of the broader developer community, it aligns with our objective to understand signing practices at a strategic level. Less experienced developers may interpret or prioritize signing differently, reflecting their vantage point within the software development lifecycle.
By targeting experts, we ensured that the findings are informed by those directly responsible for implementing and guiding signing strategies. 


\ul{Positionality Statement:}
\label{sec:positionality}
We acknowledge that our backgrounds may have influenced this study~\cite{dodgson2019reflexivity}.
The author team are cybersecurity researchers with expertise in software signing, so our expertise both enabled and influenced the protocol development and analysis.
In addition, one of the authors is a contributor to a popular signing tool.
To reduce bias resulting from specific aspects of that tool, that author had no involvement in the development of the research protocol or analysis.
However, that author informed the framing of the study and reviewed the results. 

\section{Results}

\subsection{RQ1: Signing Implementations in Practice}  \label{sec: Rq1_result}

Part of our interview protocol (\cref{tab:InterviewProtocolSummary}) ascertained how subjects use software signing in their security strategies at each stage of the \sscfm (\cref{fig:SoftwareFactoryModel}). From these responses, we identified when \sign was required by the subject's team (or organization) and mapped these instances back to the stages of the \sscfm. The summary of our framework analysis is presented in~\cref{tab:rq1}.

{
	\begin{table}[!t]
		\centering
		\caption{
			Different Signing Use Points, categorized by the stages of the \sscfm highlighted in~\cref{fig:SoftwareFactoryModel} and subjects who employ \signs at these points. We mark subjects who consider them optional for reasons of demand by customers, nature of products, team requirements, etc. The points highlighted (PI, VI, PE, VE, PS, VS, PB, VB, PA, VA, PD, VD) are described in ~\cref{fig:refined_SoftwareFactoryModel}. 
		}
		\small
            \scriptsize
		\begin{tabular}{ p{0.57\linewidth}p{0.33\linewidth}}
            \toprule
            \textbf{Points \& description of Software signing use } & \textbf{Subjects} \\
            \midrule
            \textbf{SOURCE CODE} & {} \\
             Internal code contributors sign commits-\textbf{PI} & S2-S11, S14-S17 (S4, S8, S14--Optional) \\
             External code contributors sign commits-\textbf{PE} & S7, S10 (S11, S14--Optional)\\
            Verify signatures from Internal contributors-\textbf{VI} & S7, S14, S17 \\
            Verify signatures from external contributors-\textbf{VE} & -- \\
            Signing after code reviews/audits-\textbf{PS} & S2-S11, S14-S17 (S4, S8, S14--Optional)  \\
            \midrule
            \textbf{BUILD} & {} \\
            Verify source code signatures before build-\textbf{VS} & S2, S3, S7, S14 \\
            Signing of build output-\textbf{PB} & S1-S5, S7, S8, S11-S17 (S8-Optional) \\
            \midrule
            \textbf{DEPLOYMENT/PACKAGE/ARTIFACT} & {} \\
            Verify build signature before deployment-\textbf{VB} & S7, S14 \\
            Signing of Final Software product-\textbf{PA} & All S1--S18 \\
            \midrule
            \textbf{THIRD-PARTY DEPENDENCY} & {} \\
            Signing after verification and certification of dependencies-\textbf{PD} & S6, S7, S9, S16, S17, S18 \\
            Verify signatures from external dependencies-\textbf{VD} & S9, S11, S17 (S9, S11, S17 --- Optional) \\
            \midrule
            Verification by final customers-\textbf{VA} & Difficult to determine \\
            \bottomrule
		\end{tabular}
		\label{tab:rq1}
	\end{table}
}

\subsubsection{Source Code}
First, we ascertained whether our subjects used any form of external code contributors in their software authorship process.
Out of \extContribution subjects who reported a form of collaboration with open-source or other external contributors, only \reqExternal subjects used external code contributions in the production code of projects they owned (both commercial and OSS). 
We highlight the points and manner in which subjects reported the use of software signing to create their source code.

\myparagraph{External contributors sign commits}
        Most teams who had input from external code contributors required \sign.
        Out of \reqExternal subjects (4 organizations) whose teams used external code contributions in their production code, 4 of these subjects (3 organizations) require that commits be signed. However, two of those do not require signing if the contribution is toward their open-source projects or noncritical projects. Only 2 of our subjects (2 organizations) unconditionally require signing from their external contributors.
        As \emph{\Subjecttwo} put it, \myinlinequote{I got something signed from somebody I don't know...So we do the code review, and then we have a sign-off by somebody on our team that signs off from that code}. 

\myparagraph{Internal contributors sign commits}
        Signing is typically required by teams for their internal code contributors. 
        \intSign of \subjects subjects report that their team required commit signing from internal contributors.
       \JD{NEWNOTE: The inline notation for subjects is now quite long so let's put the parentheses at the end of the sentence. Please check for other cases like this. It's fine for the ``As SX said...'' case but for the multi-subject case it's too big.} 
        Three subjects (\emph{S4}, \emph{S8}, \emph{S14}) described this as an optional requirement and largely agreed with S2's reasoning that signing alone does not establish trust in identity. Notably, these subjects' teams often rely on contributions from both internal and external contributors.

\myparagraph{Verification of signatures from contributors}
        Although code contributions are often required to be signed, the signatures are not always verified to ensure the security and provenance of the contributions.
        According to our subjects, the most commonly used security methods for assessing the security of code contributions include techniques like code reviews, security audits, vulnerability scans, and tests.
        For example, \emph{\Subjectten} states, \myinlinequote{I think everybody on the team does sign their commits, but we don't really verify it...If somebody stopped signing their commits, I don't think much would really happen.}

\myparagraph{Code is signed after reviews/audit}
        The outputs of code reviews and audits are generally required to be cryptographically signed. Subjects whose teams(or organizations) mandate signing for code contributions by internal contributors also report that all activities related to code reviews and audits must be signed as well (14 subjects).

\subsubsection{Build}
Next, we examine our subjects' responses regarding how signing is used in the build process of software products.
For most subjects, the build process and the signing of build outputs (for subjects who do so) are automated.
We expect that automation increases the use of software signing. 


\myparagraph{Verification of source code signatures before build}
        Signatures from source code are rarely verified before the build process begins. Four subjects mentioned some form of rigorous verification at this stage. For those who do not use or verify signatures, alternative security techniques are regularly employed, such as automating build pipelines with workflow definitions, using build policies, implementing reproducible builds, and utilizing code scanning tools to ensure the security of inputs to the build process.

\myparagraph{Signing of build output}
        14 subjects (11 organizations) reported signing at this stage of the software delivery process.
        10 subjects (6 organizations) reported signing alongside attestations, SBOMS, and other metadata. 


\subsubsection{Package/Artifact}
Next, we analyze the use of signing at the final stage of deploying software products.

\myparagraph{Verification of build signature before deployment of final artifact}
        Similar to the other verification steps, signatures from the build phase are rarely verified before the deployment. 
        Only two subjects (1 organization) mentioned verification at this point.
        For example, \emph{\Subjectseven} states, \myinlinequote{We sign our code and then we also verify that it was built in our specific build system.} 


\myparagraph{Signing of Final Software product}
    All \subjects subjects said that signing is done at the deployment stage to establish provenance for the organization on each of their final products. 

\subsubsection{Third-Party Dependencies}
\label{sec:third_party_deps}
For third-party dependencies in the software supply chain, we assessed how \signs influences dependency selection as a security factor and how signatures are used to verify the authenticity of dependencies in practice.

\myparagraph{Presence of a Signature as a Factor for Dependency Selection}
When they select third-party dependencies, most subjects do not consider whether it is signed. Only three subjects view signatures as an indicator that the package maintainer prioritizes security. Other security factors influencing the choice of third-party dependencies include the use of OpenSSF's scorecard tool, known vulnerabilities, level of activity, licensing, and the age of the dependency.
        
\myparagraph{Verification of Signatures from external dependencies}
        Signatures on third-party dependencies are rarely verified in practice. Only three subjects (3 organizations) reported some form of verifying signatures from third-party dependencies.
        Signatures are mainly verified for commercial off-the-shelf software.
        As \emph{\Subjecteleven} says, \myinlinequote{If it is open source, then probably no...But if it is commercial off the shelf, we want to make sure those are signed.} 
        %
        However, most of our subjects reported several methods for assessing the authenticity and integrity of a third-party dependency.
        These methods include code and vulnerability scanning (\eg Coverity), source rebuilding, internal mirroring of dependencies, and reliance on ecosystem internal checks (\eg Go package manager's checks).
        Other subjects perform internal reviews of dependencies.
        Some have a centralized dependency selection and management process, where the organization hosts and maintains dependencies  (5 subjects from 4 organizations).
        A few subjects noted that their organization discourages or forbids third-party dependencies (S13 \& S8).
        In addition to relying on alternative methods to assess dependency security, most of our subjects echoed the sentiment that OSS dependencies are rarely signed and that the identity of individuals signing open-source dependencies does not inherently establish trust.

\myparagraph{Signing after verification and certification of dependencies}
        A few subjects reported that third-party dependencies are thoroughly reviewed using one of the aforementioned techniques before being signed off for use by a team member.
        Six of our subjects indicated that this process is followed in their software production workflow.
        For example, \emph{\Subjectseventeen} described their team's process as this: \myinlinequote{We have a signed report that's signed by us, and we trust us...that says, `Hey, I scanned this third-party dependency, did all these things. I don't have a signature for that, but I ran all these extra checks and then signed those checks. So now we have a record.'} 
        In contrast to that attestation approach, \emph{\Subjecteighteen} says their team ingests third-party dependencies and signs the internal versions directly.
        

\subsubsection{Refined Software Supply Chain Factory Model}\label{refined_SSCFM}
As a result of our framework analysis, we identified various points where \signs is utilized.
We compared these to the descriptions in several Software Supply Chain security frameworks, especially the SSCFM (\cref{sec:Research_Design},~\cref{fig:SoftwareFactoryModel}).

\begin{figure*}
    \centering
    \includegraphics[width=0.65\textwidth]{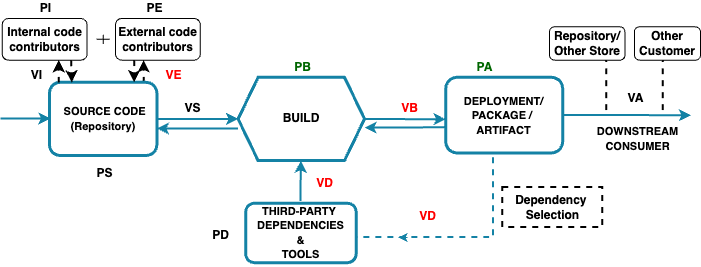}
    \caption{ 
    Our refined software supply chain factory model (compare to~\cref{fig:SoftwareFactoryModel}), highlighting different points where provenance and integrity of software could be established (PI, PE, PB, PD, PS, PA) and Verified (VI, VE, VB, VD, VS, VA) using software signatures. These points were identified from participants' responses.
    Note that while signing at PB and PA stages was commonly practiced (in green), verification at VD, VB and VE stages was minimally performed (in red).
    Abbreviations are defined in~\cref{tab:rq1}.
    }
    \label{fig:refined_SoftwareFactoryModel}
\end{figure*}

We observed that some of these frameworks and standards are ambiguous, particularly in specifying where within the \sscfm practitioners are required to establish provenance, whether by creating or verifying a signature.
We therefore present a refined version of~\cref{fig:SoftwareFactoryModel} in~\cref{fig:refined_SoftwareFactoryModel}. This updated model clearly identifies the points where provenance should (could) be established, providing guidance for organizations and teams newly adopting software signing in their software creation process. For example, our refined model is compared with the SLSA framework ~\cite{SLSA_The_Linux_Foundation_2023}. Despite SLSA's rich description of security recommendations and its focus on build levels, there is surface-level confusion about whether provenance requirements apply solely to the build stage or also to earlier stages such as collaboration.

\subsection{RQ2: Challenges Affecting the Implementation of Software Signing in Practice} \label{implem_challenges}
Our next research objective was to understand the various challenges practitioners face while implementing \signs in practice.
We grouped subjects' responses under three main themes for challenges faced by practitioners in practice; Technical, Organizational, and Human factors. 
\cref{tab:rq2} gives a summary.
Due to space limitations, we focus on only a subset of the individual challenges.

{
	\begin{table*}[!t]
		\centering
		\caption{
			Challenges to \signs implementation in practice. We have categorized related challenges highlighted by subjects. 
            Additional technical challenges, 'Managing Signed Artifacts' and 'Identity Management/Authentication,' are detailed in Appendix~\ref{sec:appendix-ExtraChallenges}, as they are not exclusive to signing tools.
		}
		\normalsize
		\small
		\begin{tabular}{lccl}
            \toprule
            \textbf{Observed Challenges} & \textbf{\#Subjects} &\textbf{\#Orgs} &  \textbf{Subjects' Proposed Solutions} \\
            \toprule
        \textbf{Technical} & {} & {} \\
        Key Management &  10 &  9 & Use of Keyless Signing (\eg Sigstore)\\
        Compatibility Issues & 6 &  6 &  ---  \\
        Lack of Verification of Signatures & 6 &  5 &  Signed Metadata, Component Data Management  \\
        Ease of Use/Usability & 4 &  4 & Usable Signing Tools (\eg Sigstore), Documentation \\
        No Unifying Standard & 2 &  2 & --- \\
        \midrule
        \textbf{Organizational} & &  \\
        Operationalization of the Signing Process & 4 &  4 & Automating Signing   \\
        Resources to Set up Signing & 3 &  3 &  ---  \\
        Creating Effective Signing Policy & 2 &  2 & Regular Process Feedback Mechanisms   \\
        No Management Incentive to Sign & 2 &  2 & ---  \\
        Bureaucracy & 1 &  1 & ---  \\
        \midrule
        \textbf{Human} & {} & {} \\
        Expertise in setting setup and use of signing & 5 &  4 & ---  \\
        Developer Attitude to Signing  & 3 &  3 &  Automating Signing   \\
        Lack of Demand from Customers & 1 &  1 &  --- \\
            \bottomrule
		\end{tabular}
		\label{tab:rq2}
	\end{table*}
}

\subsubsection{Technical Challenges}

\myparagraph{Key Management}
Key management issues were the most reported issues.
Issues under this category vary from \textit{key management, key distribution, key generation, and public key infrastructure (PKI)} issues.

\textit{Key management} issues are the highest categories recorded. \emph{S13} articulates this, especially concerning key distribution: \myinlinequote{a lot of the ways in which I think previous teams' software signing hasn't been useful is...key management...in terms of how the private key is managed but also in how the public key is made available.}


Some subjects discussed recent signing technologies that implement keyless signing (\eg Sigstore) as a possible solution.
\emph{\Subjecteighteen} said: \myinlinequote{Sigstore comes in...keyless signing...you don't have to worry about long-lasting static SSH keys, or keys being compromised, or rotating keys...you don't have these big servers to store all these signatures}. 
\emph{\Subjectseventeen} also noted, \myinlinequote{[Sigstore's] other strength...I can associate my identity with an OIDC identity as opposed to necessarily needing to generate a key and keep track of that key and yada, yada. So that's super useful because I could say, `Oh, this is signed with [a] GitHub identity.' So unless my GitHub identity has been compromised, that's much better}. 


\myparagraph{Compatibility Issues}
Compatibility issues arise during the integration of software signing systems and tools with other tools within the software engineering process, including other security tools and the artifacts to be signed, they also include scenarios where challenges with signing were linked to other security toolings,. 

\emph{S2}, \emph{S7}, \emph{S17}, and \emph{S18} highlight issues in the compatibility of newer signing systems with existing software systems in the software engineering process. 
\JD{NEWNOTE: The text says S18 in the previsou sentence, nad now this one is Subjecteight. I suggest you make all S into macros. And have both a short macro and a long one (SubjectEightLong, SubjectEightShort) --- you could list the orgs in the previous sentence, but if the detailed note is right there then you could then just write ``S8'' without the subscript in the next sentence.}
\KC{NEWNOTE:I would change all to macros. But for now for some reason, if I change it to macros now, it is not highlighted in blue}
\KC{Fixed}
As \Subjecteighteen put it, \myinlinequote{There are teams who've been doing this for quite a while...Selling that is a big challenge, right? Because we need to prove that it is much more secure than what has been there in place.}


\emph{\Subjecteight}, whose organization's software product is built for different platforms and hardware architectures, noted compatibility issues across various architectures: 
  \myinlinequote{Every platform evolved separately...macOS...Windows...The requirements are changing and evolving independently}.

Compatibility issues can also occur in situations where current security test tools are not designed to identify issues with \sign, as noted by \emph{\Subjectfive}:
\myinlinequote{The vulnerability scan report [is for]...code...CVE published...[but] signatures, I don't think it will pick those things up.}

\myparagraph{Ease of Use/Usability}
Usability of the signing tool or implementation was a common challenge hampering effective use in practice.
This was often expressed as developer experience, \eg \emph{\Subjectseventeen} said:
\myinlinequote{For a developer, make sure that they can sign the commits easily}. 
This point was also raised at a team level by \emph{\Subjectthirteen}: \myinlinequote{I think that awareness of the value of signing and sophistication around how signing is performed has increased significantly, especially in recent years. But it has also resulted in a lot of smaller teams deciding that this is not a manageable thing.} 

Concerning documentation, \emph{\Subjecteleven} said: \myinlinequote{I think there are multiple challenges there. When you build some signing infrastructure, you want to make it a self-serve so the teams can onboard themselves, how well you have your self-serve playbooks written down, documentation on using those. I think documentation can be improved.}


\myparagraph{Lack of Verification of Signatures}
Lack of verification of signatures is also quoted by several subjects as a main challenge. This arises from the factors like
  signing serves as a ``checkmark'' security tool (\Subjectfour: \myinlinequote{most people who are strictly signing all commits don't have a meaningful process for verifying those signatures or for enforcing any policy around them. They just do it because}),
  and
  lack of customer interest (\Subjectfourteen: \myinlinequote{We have tried to ... get potential customers and customers interested...but as far as I know, none of our customers...verify our signatures on images.})

\myparagraph{Verification of Artifact Before Signing}
Verification generally was highlighted as a problem, especially verifying inputs to the build process of the software delivery process.
The problem noted here is that the content of software artifacts is not readily verified before signing: as \emph{\Subjectsixteen} put it, \myinlinequote{if you start with the component, the component gets put into a subassembly. The subassembly gets put into a package. The package can get put into a broader package or whatever. I think the people are signing. But do you know that the sum of all the pieces are accurate? ... what are you really signing, and at what level?}

\subsubsection{Organizational Challenges}

\myparagraph{Operationalization of the Signing process}
The implementations of \signs are typically neither uniform nor centralized across different organizations. Various teams are largely responsible for crafting their own solutions. 
As \emph{\Subjectsix} said, \myinlinequote{I can't be more specific, but today, I've talked to three different teams with three different processes}. 
%
%
\emph{\Subjectfifteen} suggested that automation could improve uniformity: \myinlinequote{We try to break down the barriers with really good examples, by having common configuration of CI through similar GitHub workflows, and stuff like that. So it's easier for people to use.} 

\myparagraph{Resource to Set up Signing}
The issue of organizational resources to set up efficient signing tooling across an organization was also another cause for concern for some of our subjects. Subjects discussed this in terms of finances, time, and engineering effort. 
\emph{\Subjectfourteen} said, \myinlinequote{We need to do a lot of signing...hundreds of container images [daily]...It was a pretty significant engineering effort to build a pipeline that can handle all those signatures.}



\myparagraph{Creating Effective Signing Policy}
Policies surrounding the implementation of software signing by teams are generally not considered highly effective.
These policies include elements such as role-based access, key usage, and key issuance. 
\ParticipantQuote{\Subjectthree}{... the other thing is policies, creating a policy that is, for one, it's hard to do. And then the other part is creating an effective policy and knowing that it's effective.}

A suggested solution to address this issue is the use of regular feedback mechanisms to assess the security policies and processes of different teams within the organization. 
As \emph{\Subjecteleven} said, \myinlinequote{We want to be more proactive...internal audits, and an assessment methodology like the NIST, I think CISF (NIST's CSF).} 

\myparagraph{No Management Incentive to Sign}
A lack of management incentives was also a challenge. 
\Subjecteleven said, \myinlinequote{The investment, the time that the engineer is putting into...signing...before a release, does it get the right attention from the senior management? Do they view it as a time well spent for the product?}

\subsubsection{Human Challenges}

\myparagraph{Expertise}
Some subjects noted that setting up signing tools depends on the engineer's expertise.
Subjects felt that most engineers lack the necessary expertise.
As \emph{\Subjectnine} said, \myinlinequote{there's a lift there as far as...expertise}.

\myparagraph{Developer Attitude to Signing}
Software developers' attitudes toward \signs pose challenges to the signing process. Disinterest and resistance to adopting new tools are the primary attitude-related obstacles cited.
\emph{\Subjectten} observed: \myinlinequote{Most developers don't really understand this stuff [or] want to think about it...[they are] happy if they...see a big green tick[mark].}
\emph{\Subjectsix} added: \myinlinequote{They refuse to get off their old tools.}

\myparagraph{Lack of Demand from Customers}
Most customer-driven organizations report that \signs is challenging to implement because customers rarely request this feature in their requirements.
\emph{\Subjectfourteen} put it like this: \myinlinequote{Sign[ing] our images was there from day...we haven't found many customers that are too excited about the signing aspect.} 





\subsection{RQ3: Perceived Importance of Software Signing in Mitigating Risks} \label{sec:RQ3}
In this section, we aimed to understand how practitioners 
perceived the importance of software signing in their software 
delivery security. Opinions on the significance of signing in the 
software delivery process varied. Subjects identified three distinct 
perspectives on the importance of software signing, reflecting on
their personal views and how signing was treated in their current 
and previous roles.

\subsubsection{Signing as Crucial to Provenance and Integrity} 
Many of our subjects considered provenance an integral part of security, and felt that software signing is (or will be) a guarantee of that. 10 subjects (representing 9 organizations) had the opinion that \sign on its own is strong enough to guarantee provenance.
\emph{S4} said, \myinlinequote{I think signing is massively important. Signing for open source is probably the single most important thing for software supply chain security}. \emph{S16}'s response mirrors this, \myinlinequote{I think [Signing's importance] is very high...There's a senior executive VP who has to attest that the software we release is what we say it is ... he could be legally sued if the software we deliver... I mean personally, not just our organization. Him, because of his level in the organization, if we publish software that we have incorrectly attested to, he can be ... So it's a very high priority.}

\subsubsection{Signing as a Secondary Security Technique}
Some subjects believe that software signing is a supplementary method used to ensure the integrity of primary security techniques. 7 subjects (6 organizations) had the opinion that \sign does not provide enough guarantee for provenance.
They often deemed it important only in conjunction with methods like Attestations and other Manifest forms such as the SBOM. For these organizations, signing is a valuable part of a broader, multi-layered security strategy.
 Some subjects who view signing as a strong guarantee of provenance also emphasized that adding a manifest or attestation further strengthens this guarantee.
As \emph{\Subjecttwo} said, \myinlinequote{I think the most important thing is accurate metadata collection and centralization. And even if that data's not signed, we can still get a lot of usefulness out of it. Signatures, in my opinion, they offer an additional layer of security on top of what we should be doing, which is metadata collection.}
\emph{\Subjectseventeen} similarly views signing as a final check on top of other security practices: \myinlinequote{Signing was essentially the way of saying, `I have checked, we did all the [security practices]'.}

\subsubsection{As a Requirement-Driven Practice} 
Subjects in this category indicated that software signing was primarily used as a "check the box" technique to satisfy regulatory, framework, customer or organizational requirements. 
They did not view signing as a primary security measure. Instead, it was employed mainly to meet regulatory or contractual obligations, rather than being treated as an essential part of their security strategy. For these subjects, signing was often seen as a procedural necessity—for instance, to submit a pull request, initiate a build, or make a commit—rather than as a measure to enhance security.
4 subjects representing 3 organizations fall into this category.
\ParticipantQuote{\Subjectseventeen}{... at certain organizations it felt like it (\signs) was more of a check-the-box thing of like, "Hey, we should sign it because that's what people tell us to do}
\ParticipantQuote{\Subjecteight}{...for signed binaries and code signing ...a lot of those [requirements] are driven...by platform requirements...to prevent us from being blocked from...these platforms, code signing...was largely driven by [these platforms].}


\subsection{RQ4: Impact of Internal \& External Events on Software Signing Practices}
\label{rq4}

In this section, we explore the impact of two kinds of security events:
  security failures such as Log4j,
  and
  the advent of security standards and regulations:
\vspace{0.08cm}
\noindent
\begin{enumerate}
    \item \textbf{\textit{Experienced \ssc Failures}}: We consider these internal if a subject's organization is a direct victim of a \ssc attack,
    and external if a subject's organization is not directly impacted in a \ssc event, \eg a general \ssc incident like the Solarwinds attack. 
    \item \textbf{\textit{\ssc Standards}}: These are industry wide external events \eg the publishing of the EO-14028 executive order on \ssc security. 
\end{enumerate}
We asked subjects about their organizational experiences with \sscsm failures (attacks, vulnerabilities, and other incidents) and how these experiences affected their team's or organization's \sscsm security decisions, particularly regarding software signing. Additionally, we queried subjects on the impact of \sscsm security frameworks, standards, and regulations on their use of \signs and other security techniques.

In their responses, subjects described a security failure they had experienced.
We categorized each failure as a Vulnerability (unexploited security flaws), an Incident (SSC failures with no known malicious intent), or an Attack (deliberate and malicious failures).
Then, subjects described changes these failures prompted in the organization.
We categorized these as Direct Fixes (patched but no change to organizational security approach), General Security Process (large-scale changes involving multiple security techniques), and Signing (adoption or change in the implementation of \signs).

\subsubsection{Impact of Experiences with SSC Failures}
\label{rq4a}

{
	\begin{table*}[!t]
		\centering
		\caption{
            Kinds of \sscsm failures experienced, and associated security changes.
		 Only one participant reported an influence of these failures on their software signing implementation. Software supply chain failures were more likely to prompt a direct fix rather than a change in the security process.	
            }
		\normalsize
            \small
		\begin{tabular}{lcccc}
            \toprule
            \textbf{Type} & \textbf{\# Subjects (\# Orgs)} & \textbf{\# Direct fix} & \textbf{\# Change in General Security process} & \textbf{\# Change in Signing process} \\
            \toprule
            Vulnerability & 9 (6) & 9 & 2 & 1 \\
            Incident & 5 (5) & 5 & 2 & 0 \\
            Attack & 4 (4) & 4 & 2 & 0 \\
            \bottomrule
		\end{tabular}
		\label{tab:rq4}
	\end{table*}
}


\cref{tab:rq4} summarizes these events and the resulting changes.
Most cases were non-malicious, either Vulnerabilities or Incidents.
However, four subjects had experienced attacks, with one major attack leading to extensive collateral damage and a complete overhaul of software and security processes. 

Across these events, the typical resolution was via Direct Fix, without an infrastructural change in the general security process or the signing process specifically.
When they occurred, changes to the general security process typically involved adoption of a group of tooling such as for SBOMs.
As a result of one event, a vulnerability led to changes in the signing process.
In this case, the team (which had a safety-critical application) integrated in-toto’s signed attestation capability.
The subject \emph{\Subjecttwo} stated: \myinlinequote{We saw that [in-toto] allowed you to decouple the security metadata collection and validation...that
allowed us to solve this problem of how do we know that this opaque process for the software, that opaque certification...was followed without...knowing the details of...that process.}

\subsubsection{Impact of Security Regulations, Frameworks, and Standards}
\label{rq4b}

Our subjects reported that security frameworks, standards, and regulations had \textit{no direct impact} on their organizations' adoption of software signing as a security method. However, these standards and regulations did influence the general adoption of other security techniques, such as SBOM. 9 participants from 5 organizations shared this view, while only 2 (from 2 organizations) from this group explicitly linked this to signing. 5 participants (from 4 organizations) provided unclear statements (personal opinions or unfamiliarity with how decisions are made).


We asked our subjects, ``Are [your SSC security] strategies influenced by regulations and standards?''. 
\emph{\Subjectone}'s response was typical: 
\myinlinequote{Yeah, I would say certainly the collection of SBOMs and doing that more broadly has been a response to things like the executive order, at least for our team specifically...The executive order played just a strong role in what we are starting to do now. Whether some teams are now trying to be preventative, that's a good question.}

\section{Discussion} \label{sec:Discussion}

\subsection{Impact of Organizational Size and Type of Produced Software on Signing Practices}
\JD{NEWNOTE: The first sentence is not interesting, just cut it and get to the point. ``We did not observe any change in XXX (RQX, Figure YYY). However, we did observe some effect in the challenges noted (RQX) and in the relevance of standards and regulations (RQY).''}
\KC{NEWNOTE: fixed}
We did not observe any significant effect of organizational size (and nature of software product) on where signing was implemented (RQ1, ~\cref{fig:refined_SoftwareFactoryModel}). However, we did note patterns in the challenges reported (RQ2) and the influence of standards and regulations on the adoption of \signs (RQ4).
It is important to note that these observations represent correlations rather than causations. While our findings highlight potential relationships, further study is necessary to establish causative factors and better understand the mechanisms driving these patterns.

\vspace{1em}
\noindent{\bf Effect on Challenges.}\;
\JD{NEWNOTE: Please list all effects.}
\JD{NEWNOTE: Macros here please.}
\JD{NEWNOTE: The topic sentence says the differences are attributable only to organization size, but the two example include (second one) that it's product type. You can either make two paragraphs (probably better) or one big pararaph with a more general topic sentence (please don't)}
\KC{fixed}
We observed a noticeable difference in the organization size of participants who reported certain type of challenges.
Only subjects from large organizations (S6, S8, S13, S15) talked about difficulties in operationalizing the signing process. 
We conjecture this is because of the greater coordination effort required to harmonize the engineering process of several teams.

The nature of the software product also seem to exert some influence on the challenge experienced.
Unsurprisingly, only subjects from organizations with non-security-focused product areas (\Subjecteleven and \Subjectthirteen) reported a lack of management incentive as a barrier to signing. 
While Only participants from security-focused product teams (\Subjectthree, \Subjectten) reported challenges in creating effective signing policies.

\vspace{1em}
\noindent{\bf Effect on Influence of Standards and Regulations.}\;
As we reported, nine participants from five organizations noted that standards and regulations influenced the adoption of tools like SBOM. Within this group, only two participants (from two of the five organizations) explicitly linked this to signing. Notably, all five organizations were either large or focused on security-related products.

\subsection{Relation between Participants' perceived importance of Signing and Signing Implementation }

We remark on a recurring dissonance between the perceived importance of \signs by the participants and their implementation practices. Although all organizations signed their final products, signatures were rarely used to verify integrity or provenance \cref{tab:rq1}. Ten subjects (from nine organizations) regarded \signs as a very strong guarantee of provenance, but only three participants (from two organizations) reported verifying signatures for internal contributors, and none did so for external contributors. 
This suggests a unidirectional use of \signs, focusing on establishing trust for their outputs while placing lower expectations on the provenance of their inputs. As a result, these organizations are effectively publishing software with a higher degree of provenance than they demand from their dependencies or external contributions. 

\subsection{Recommendations from this study}

\noindent{\bf Improving Signature \& Artifacts Verification in Signing Workflows.} \SO{just another reminder to check your capitalization for your headings because it is not consistent}
\; Our refined software supply chain factory model has highlighted a significant gap \SO{"highlighted a significant gap" I think the use of the word "gap" is off and you could be more intentional with what you want to say for this first sentence. Maybe along the lines of "Our software supply chain factory model analysis has revealed significant shortcomings in the usability and implementation of software signing.} in the current practice of signature verification. This issue was underscored by several subjects in our study.

In certain automated workflows, the act of signing is often treated as a mere formality, serving as a ``checkbox'' that, once done, allows engineers to carry out their usual tasks (e.g., creating pull requests or merging code into the main branch). Thus, this checkbox \SO{If the "checkbox" is software signing" I don't think it's safe to say in your argument that the task itself doesn't provide security properties.} provides very little security properties, and requires effort to provide meaningful security gains.

To strengthen provenance guarantees, signing should be complemented with transparency-based methods, ensuring both the signer’s identity and artifact integrity. This requires consistent verification of signatures and artifact content to provide meaningful security benefits.


Future works in this area should focus on proposing improved tooling around existing or new software signing tools to include these capabilities in their design. This enhancement would not only improve transparency but also reduce the costs and expertise requirements for setting up such infrastructure.
Also, while much of our recommendations focus on ``improving tooling'', further research is needed to define what kinds of tooling would qualify.
\JD{NEWNOTE: Cite the Taylor magazine article (arXiv version) here.}
\KC{added}
This research might involve experimental and empirical analyses of the current generation of software signing tools, particularly in terms of functionality, integration~\cite{schorlemmer2025establishing}, and usability. 

\JD{NEWNOTE: This is a pretty weak topic sentence}
\KC{NEWNOTE: fixed}
As additional takeaways, we emphasize the need for further research to address critical trust gaps within the open-source ecosystem. Our findings (~\cref{sec:third_party_deps}) highlight an underlying lack of trust in open-source packages, suggesting that cross-ecosystem provenance maintenance could serve as a potential solution.

By addressing these issues, we can ensure that signing is not just a procedural element and instead plays a crucial role in maintaining the integrity and authenticity of the software development process.

\vspace{0.9em}
\noindent{\bf Addressing Incentive Challenges in Software Signing.}\;
One of the most frequently reported challenges in the realm of software signing is the lack of proper incentives for its adoption. These incentives can range from support from management, the ease of using signing tools, as well as comprehensive education on the importance and application of signing.
Although some of these issues, such as the need for educating practitioners on the significance and application of signing, can be addressed with relative ease, others necessitate more elaborate planning. For instance, fostering management support might require demonstrating the tangible benefits of software signing in terms of enhanced security and reduced risk of software tampering. 
Moreover, though the user-friendliness of signing software has been evidenced by various ``Why Johnny Can't'' papers, more research is needed in lowering the adoption bar for software signing systems. 
Regardless of the complexity of these challenges, investing in these areas is paramount for enhancing an organization’s security posture. By fostering a culture that values and understands the importance of software signing, organizations can ensure the integrity of their software and protect against potential security threats.

Future research could focus on developing strategies to address these incentive-related challenges and exploring their effectiveness in different organizational contexts. This could provide valuable insights for organizations striving to improve their software signing practices.

\subsection{Contrast to prior work}

%

\vspace{0.9em}
\noindent{\bf Comparisons with Previous Quantitative Studies on Software Signing.}\;
We compare our findings with Schorlemmer \etals ~\cite{schorlemmer_signing_2024} quantitative study on software signing in open-source ecosystems. Their results indicate that major software supply chain failures (e.g., SolarWinds, Codecov attacks) and dedicated tooling had no significant impact on the quantity of signatures. In contrast, our findings for the non-open source industry show the opposite. This difference is expected due to the substantial incentives, such as commercial and reputation losses, that drive greater adoption in the non-open source industry. Additionally, our study supports this finding, as some subjects noted that contributions to the open-source ecosystem often involve relaxed security requirements.

\vspace{0.9em}
\noindent{\bf Distinctions of Refined Model Compared to Existing Frameworks.}\;
While all software supply chain security frameworks agree on the defined threat model, our refinement of this model differs from existing ones in several ways.
\begin{enumerate}
    \item SLSA: The SLSA framework primarily focuses on securing the build process within the software engineering lifecycle, offering recommendations as a series of incremental security hardening levels. In contrast, our refinement addresses the entire software supply chain, not just the build step.
    \item SCIM and S2C2F: This framework provides threat-reduction recommendations without specific implementation details. Our approach, however, emphasizes the continuous establishment and verification of provenance across the software supply chain.
    \item Hammi \etal ~\cite{Hammi_2023}: While their work proposes a blockchain-based cryptographic tool, we do not advocate for a specific tooling solution. Instead, our model is designed for industry and software teams to create an implementation workflow that is independent of specific tools. Additionally, our findings are supported by insights from industry practitioners, enhancing the practical applicability of our model.
\end{enumerate}

\section{Conclusion}

\JD{NEWNOTE: I think we will want to make some minor edits here and in the Abstract and in the Intro and the Background to address misc. comments from the reviewers.}

Software signing is widely recommended for ensuring software provenance, yet many organizations face challenges when integrating it into their processes. In this first qualitative study, we examine the practices and challenges of software signing through interviews with 18 senior practitioners across 13 organizations. We identified that while software products are signed, the signatures often go unvalidated. Key barriers include infrastructure limitations, lack of expertise, and insufficient resources. Although signing offers strong provenance guarantees, it remains secondary in many organizations' cybersecurity strategies. Additionally, organizational behaviors are influenced by major incidents like SolarWinds and industry regulations and standards, with concerns raised about the quality and development of these standards.

Our findings shed light on the real-world practices, challenges, and organizational factors shaping software signing adoption, offering guidance for organizations on their journey toward more secure software supply chains.

\ifANONYMOUS
\else
\section{Acknowledgments}
We thank the study participants for contributing their time to this work.
Taylor Schorlemmer, Sofia Okorafor, and Wenxin Jiang assisted in typesetting.
We acknowledge support from Google, Cisco, and NSF \#2229703.
\fi

\raggedbottom


\section{Ethical Considerations}

This work considers the human and organizational factors governing the adoption of software signing techniques in software engineering work.
To elicit this data, we developed a research protocol, including subject recruitment, interview content, and data processing and analysis.
This protocol was approved by our organization's Institutional Review Board (IRB) prior to the study conduct.
All members of the research team who interacted with subjects or their non-anonymized data have completed our organization's training procedures for handling personally identifiable information.
Data from individual subjects were anonymized following best practices for qualitative data analysis.

Any human-subjects research effort must weigh the risks and benefits for individual research subjects as well as for society.
The primary \textit{risk} for subjects is the hazard to their individual or organizational reputation when describing what might be interpreted as sub-standard cybersecurity practices.
We therefore anonymized subjects by rank and by organization.
A secondary cost for subjects was their time.
Subjects were not coerced to participate, and we offered an incentive of \$100 for their participation.
The primary potential \textit{benefit} for our subjects, their organizations, and the societies in which they operate, is that the resulting data may motivate improvements in the software signing ecosystem, as we discussed in~\cref{sec:Discussion}.
In our judgment, the potential benefit to software cybersecurity far outweighs the risks posed to our subjects and their organizations.

We therefore attest that:
\begin{itemize}
    \item The research team considered the ethics of this research, that the authors believe the research was done ethically, and that the team's next-step plans (\eg after publication) are ethical.
 \end{itemize}

\section{Compliance with Open Science Policy} \label{sec:OpenScience}
We acknowledge the commitment to open science, where authors are generally expected to share their research artifacts openly. The research artifacts associated with this study are
\begin{itemize}
    \item Raw transcripts of interviews
    \item Anonymized transcripts of interviews
    \item Interview protocol
    \item Codebook
\end{itemize}

\myparagraph{Things we have shared}
The \textit{interview protocol} is a crucial part of any interview study, since it allows for the critical review of a study design as well as its replication.
Our full interview protocol is included in~\cref{sec:appendix-InterviewProtocol}.
Since it is semi-structured, we include all of the questions asked of all subjects, as well as examples of the follow-up questions we asked.
We also share the \textit{codebook}, with codes, definitions, and example quotes that we coded for each code --- see~\cref{sec:appendix-Codebooks}.

Our full \textit{codebook} is available at \url{https://doi.org/10.5281/zenodo.14660193}.

\myparagraph{Things we cannot share}
For subject privacy reasons, and for IRB compliance, we cannot share the raw transcripts.
We are also unwilling to share the anonymized transcripts of the interviews.
Given the high organizational ranks of many of our subjects, and the small size of the subject pool resulting from our recruiting strategy, we believe there is a high risk of de-anonymization even of anonymized transcripts. Therefore, we cannot share the anonymized transcripts.

\raggedbottom
\pagebreak

\clearpage


\bibliographystyle{plain}
\bibliography{bibliography/references, bibliography/new}

\appendix

\section*{Outline of Appendices}

\noindent
The appendix contains the following material:

\begin{itemize}[leftmargin=12pt, rightmargin=5pt]

\item \cref{sec:appendix-signing}: Traditional Software Signing Workflow.
\item \cref{sec:appendix-Demographics}: Organizational Demographics.
\item \cref{sec:appendix-CodeSaturationCurve}: Code Saturation Curve.
\item \cref{sec:appendix-ExtraChallenges}: Uncommon (less reported) technical challenges, omitted from the main paper due to space constraints and their broader applicability to tools beyond signing.
\item \cref{sec:appendix-InterviewProtocol}: The interview protocol.

\item \cref{sec:appendix-Codebooks}: The codebooks used in our analysis, with illustrative quotes mapped to each code.

\item \cref{sec:appendix-CodeEvolutionDiagram}: Codebook Evolution Diagram 
\end{itemize}

\section{Traditional Software Signing Workflow} \label{sec:appendix-signing}

\begin{figure}[htbp]
    \centering
    \includegraphics[width=0.95\linewidth]{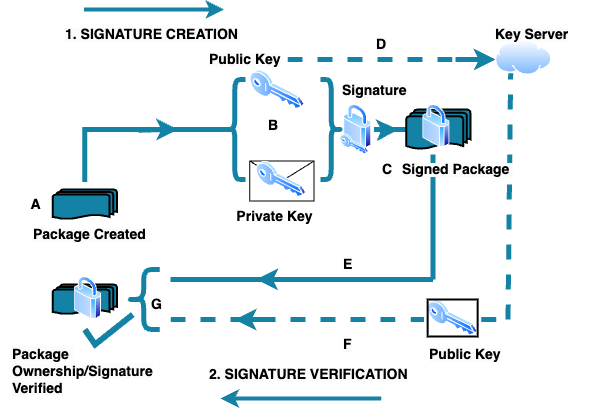}
    \caption{ 
    Typical workflow for software signing and verifying signatures.
    The component author packages (A) and signs (B) their software.
    The signed package (C) and public key (D) are published.
    To use a package, a user downloads it (E) and its public key (F) and verifies the signature (G).
    }
    \label{fig: Signing-fig}
\end{figure}

We show a typical traditional package signing workflow in~\cref{fig: Signing-fig}.

\section{Organizational Demographics} \label{sec:appendix-Demographics}
We summarize additional organizational demographic information of our subjects in \cref{tab:orgSummary}. This table provides further context for the subject's organizational size, type of software product, and how many of our subjects belonged to each organization.

{
\renewcommand{\arraystretch}{1.3}
\begin{table}
    \centering
    \caption{
    Organizational Demographics. To enhance anonymity, We only highlight the number of subjects in each organizational category. letters A-L is used to depict each organization.
   }
   \small
    \scriptsize
    \begin{tabular}{
    p{0.3\linewidth}p{0.6\linewidth}
    }
    \toprule
        \textbf{Type}& \textbf{Breakdown (\#Organizations|\#Subjects)} \\
        \toprule
         Organizational Size (Employee Size)& {
            Small (<100) (4/6), Medium (<1500) (3/4), Large (>1500) (6/8)
         } \\
         Product Area & {
             Digital technology (1/3), SSC Security (2/4), Social technology (1/1), Dev tools (1/1), Telecommunications (1/1), Cloud security (2/3), Aerospace Security (1/1), Internet services (2/2), Cloud + OSS Security (1/1), Cloud + Dev toools (1/1)
         } \\
         \midrule
        Subject Distribution& {
             A (3), B (3), C (1), D (1), E (2), F (1), G (1), H (1), I (1), J (1), J (1), K (1), L (1)
         }  \\
        \bottomrule
    \end{tabular}
    \label{tab:orgSummary}
\end{table}
}

\section{Code Saturation Curve} \label{sec:appendix-CodeSaturationCurve}

\begin{figure}
    \centering
    \includegraphics[width=0.90\linewidth]{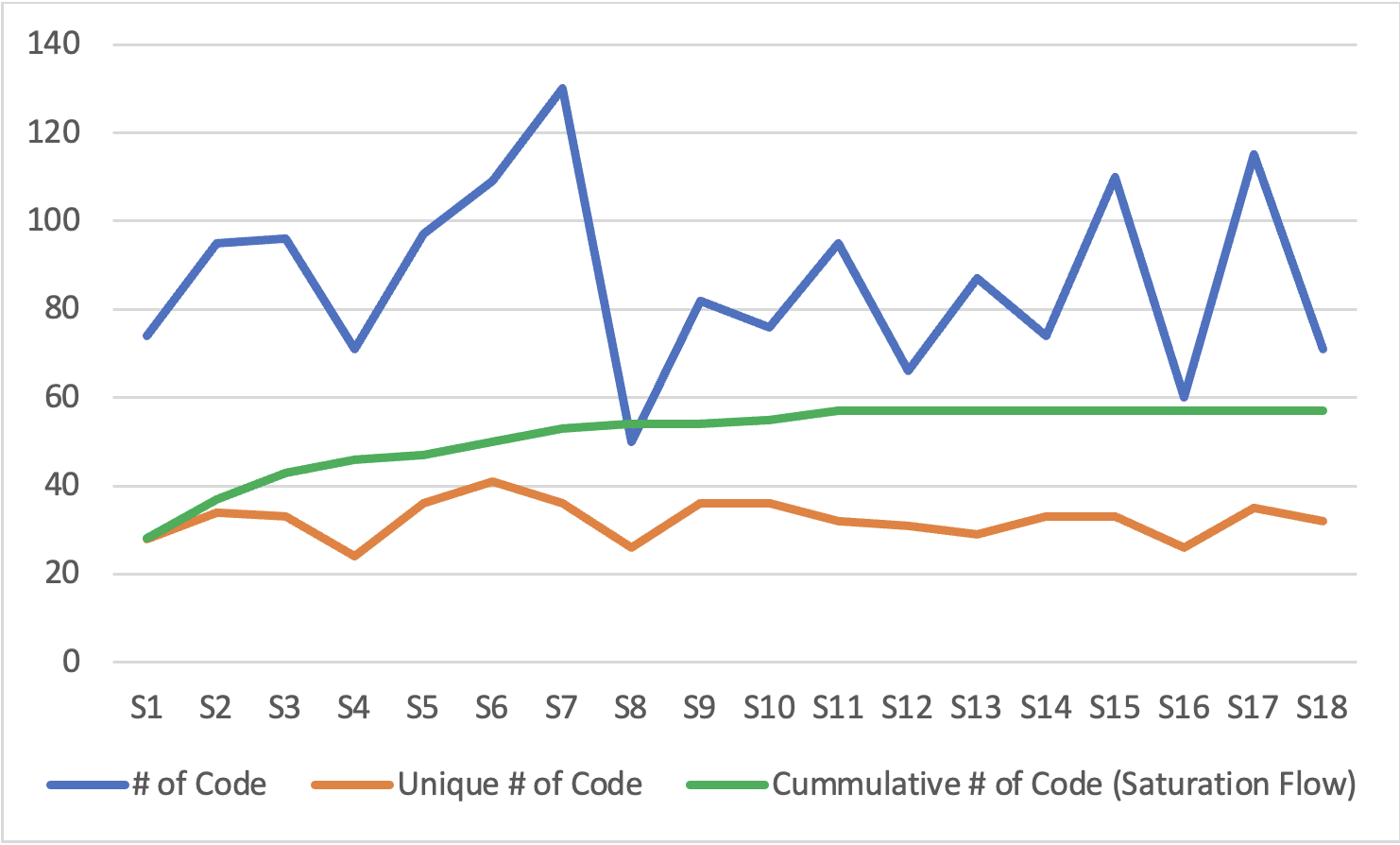}
    \caption{ 
    Saturation curve.
    Interviews are plotted in the order in which they were conducted.
    Each interview covered many topics in detail (orange line, blue line).
    However, as the green line shows, our results saturated (\ie stopped observing new perspectives) around interview 11.
     }
    \label{fig: saturation}
\end{figure}

We present the trends from the data analysis of our interview subjects in \cref{fig: saturation}. The total number of codes per interview is shown by the blue trend line, with a median of 84.5 codes per interview. Participant S8 had the fewest codes (50), while S7 had the most (130).

The orange trend line represents the unique number of codes in each interview, with a median of 33 unique codes. Subject S4 had the fewest unique codes (24).

The green trend line illustrates the saturation curve for the study, indicating that no new unique codes were observed beyond interview 11 (S11).

As noted earlier (\cref{sec:Data_Collection}), the pilot interviews (S1 and S2) were included due to minimal changes in the interview protocol. These interviews also contained more unique and total codes than the study's median values, further justifying their inclusion.

\section{Other Less Reported Technical Challenges} \label{sec:appendix-ExtraChallenges}

\subsubsection{Managing Signed Artifacts}
Managing signed software artifacts post-signing should follow good practices, but this is currently a challenge.
\ParticipantQuote{S9}{Also managing all the artifacts that are signed ... It's great that you sign things, but what do you do with it afterwards? I think that's kind of something that we're facing right now, especially when it comes to container management. Great, we signed containers, but now what? ... Okay, great, now what? All those downstream things that we're trying to figure out.}

\subsubsection{Identity Management/Authentication}
Managing and authenticating the identities of team members authorized to sign at various stages of the software delivery process poses challenges in some instances.

\ParticipantQuote{S1}{in general one of the challenges that I hear from other security architects and security leads, and program managers, and people at that level, is key management and identity management as being a barrier for signing and security and authentication in general.}

\noindent
Possible solutions suggested and used by some participants include the use of short-lived or constant key rotation. 
\ParticipantQuote{S7}{We really rely on short-lived tokens for authentication. I think that is one way that we really try to restrict access to our system and make sure that there's no long-lived secrets, there's no long-lived access tokens anywhere in our system. For any signing that we do, I think those keys are rotated quite frequently. So, that's another thing we do to prevent any attack should key information be leaked. I'm trying to think what else.}

\section{Interview Protocol} \label{sec:appendix-InterviewProtocol}

\cref{tab:InterviewProtocolSummary} gave a summary of the interview protocol.
Here we describe the full protocol.
We indicate the structured questions (asked of all users) and examples of follow-up questions posed in the semi-structured style. Given the nature of a semi-structured interview, the questions may not be asked exactly as written or in the same sequence but may be adjusted depending on the flow of the conversation.

We include Section D of the protocol for completeness, although it was not analyzed in this study.

{
\begin{table*}
\centering
\caption{
    Our full interview protocol.
    The table also indicates the history of the protocol development: removed (strikethrough), changed ($\star$) and re-arranged ($\dagger$) questions.
    We did not analyze part D in this study. We excluded this data due to limited detail and confidentiality concerns. 
    Following qualitative reporting best practices ~\cite{voils2007or}, we disclose this topic to acknowledge its potential influence on data collection.
    }
\label{table: InterviewProtocol}
\tiny
\begin{tabular}{p{0.2cm} p{1cm} p{15cm}}
\toprule
\textbf{Themes} & \textbf{Sub-themes} & \textbf{Questions}\\
\toprule
\multirow{7}{*}{{\rotatebox[origin=c]{90}{\textbf{A: Demographic}}}}
& &
\textbf{A-1:} \textit{What best describes your role in your team? (Security
engineer, Infrastructure, software engineer, etc)}\\
\\
& &\textbf{A-2:} \textit{What is your seniority level? How many years of
experience?}\\
\\
& &\textbf{A-3:} \textit{What is the team size?}\\
\\
& & \textbf{A-4:} $\dagger$ What are the team’s major software products/artifacts?\\
& & \st{\textbf{A-5:} What type of organization/company?}
\\
\midrule

\multirow{10}{*}{{\rotatebox[origin=c]{90}{\parbox{3cm}{\centering \textbf{B: Software Supply Chain}\\\textbf{Failure Experienced}}}}}

& & 
\textbf{B-1:} \textit{Describe briefly the team’s process from project conception to product release and maintenance – This is
to understand the unique context of each practitioner’s software production case.}\\
\\
& & \textbf{B-2:}  $\star$ What do you consider a Software Supply chain attack/incident to be?\\
\\
& & \textbf{B-3:} Can you describe any specific software supply chain risks (or incidences with third-party dependencies, code contributors, open source, etc.) that your team has encountered during your software development process?
    \begin{itemize}
        \item How were these addressed? (Alternatively -- How did this affect the decision to implement \signs? 
    \end{itemize}
    \\
& &\textbf{B-4:}  $\dagger$What are the team’s major software products/artifacts?
(Moved To Demographic from Prevalent SSC risk section After Pilot)\\
\\
& & \textbf{B-5:} What is your team’s greatest source of \sscsm security risk?
    \begin{itemize}
        \item Project components(Third-party dependencies, build process, code contributors, etc) 
        \item $\dagger$ Between the \sep vs project components which constitute a greater source of risks?
    \end{itemize}
    \\
\midrule

\multirow{35}{*}{{\rotatebox[origin=c]{90}{ \textbf{C: Software Signing in Mitigating Software
Supply Chain Risks}}}}

& &
\textbf{C-1:} If no incident has been recorded (depending on the answer from Section B): why did the team choose to implement software signing?
    \begin{itemize}
        \item Is software signing the team’s major strategy to secure its supply chain? Any other complimentary security efforts and methods?
        \item Are these strategies (\signs and complementary security techniques) influenced by regulations and standards?
        \item What challenges or obstacles did your team face while integrating Software Signing into your supply chain security practices?
        \item $\star$ How do different team members contribute to the implementation of Software Signing? What roles and responsibilities are involved? (After Practise).
        \item Possible Follow up -- Do you think Software Signing on its merit (if implemented right) is good enough to completely secure a supply chain? (Added after Pilot).
    \end{itemize}
\\
& \textbf{Third-party dependencies}
& \textbf{C-2:} What are the team’s peculiar selection strategies for Third-party dependencies?
    \begin{itemize}
        \item How does the presence of a signature influence this decision?
        \item How is the authenticity of this signature verified?
        \item Any other methods/practices to ensure the trustworthiness of the third-party dependencies before integrating them into your projects?
        \item Possible Follow-up -- Can you describe these methods/practices?
    \end{itemize}
\\
& & \textbf{C-3:} What influences the team to discontinue the use of a package?
    \begin{itemize}
        \item How does the Signature on the dependency influence this decision?
    \end{itemize}
    \\
\\
& & \textbf{C-4:} How does the team manage its third-party dependencies’ security?
    \begin{itemize}
        \item How do you maintain awareness of potential vulnerabilities or threats related to third-party components?
    \end{itemize}
\\
& \textbf{First-party source code} & \textbf{C-5:} Is software signing a requirement for team developers (insider threats) and open-source contributors (if any)?
    \begin{itemize}
        \item how does the team use software signing to protect its source code (what parts of the process is signing required e.g. Commit signing)?
    \end{itemize}
    \\
\\
& \textbf{Build Process} & \textbf{C-6:} (How) Does the team utilize signing in its build process?
    \\
\\

& \textbf{Package Artifact/Deployment} & \textbf{C-7:} How is Signing used to protect the following from compromise?
    \begin{itemize}
        \item how Artifact’s build binaries/deployment pipeline
        \item Artifact’s repository
    \end{itemize}
    \\
\\

& & \textbf{C-8:} $\star$ How does the team evaluate the effectiveness of their Security processes/Signing Implementations? Any Feedback mechanism? (This question is derived from  ~\cite{kalu_reflecting_2023}).
    \\
\midrule
\multirow{13}{*}{{\rotatebox[origin=c]{90}{ \textbf{D: Signing Tool Selection}}}}
& &
\textbf{D-1:} \textit{What \signs tooling does the team use (If not mentioned yet)}\\
\\
& &\textbf{D-2:} \textit{What factors did the team consider before adopting this tooling over others?}\\
\\
& &\textbf{D-3:} \textit{What was the team’s previous signing practice before the introduction of current tooling?}\\
\\
& &\textbf{D-4:} How does your team implement this tooling (which components of tool does the team use)?\\
& &\textbf{D-5:} Have you encountered any challenge(s) using this tool of choice? How have you coped with this limitation(s)?
    \begin{itemize}
        \item Are there any areas where you believe further enhancements could be made in your current implementation of these methods?
    \end{itemize}
\\
& &\textbf{D-6:} Have you/your team considered switching this tool for another?
    \begin{itemize}
        \item What other tools have been considered or are currently being considered?
    \end{itemize}
\\
\bottomrule
\end{tabular}
\end{table*}
}

\section{Codebooks} \label{sec:appendix-Codebooks}

\cref{tab:codebook} describes our codebook, with code definition, and example quote tagged with that code.

{
	\begin{table*}[!t]
		\centering
		\caption{
   Excerpts from the final codebook we used for our analysis.
   Sample codes under each theme give examples.
   Note that the theme, ``Others and Specific tooling'' is not discussed in the paper.
		}
		\scriptsize
		\begin{tabular}{p{2.0in}p{2.0in}p{2.5in}}\hline
            \toprule
             \textbf{Code} & \textbf{Definition}& \textbf{Sample Quote}  \\
            \midrule
            \textbf{BUILD} & {}  \\
            Build Infrastructure of team & Nature of Subject's build process or build Infrastructure. Example- when subject mentions that their build or CICD is automated & S7: "Our build system is typically GitHub actions."  \\
             & &\\
            \textbf{SOURCE CODE} & {} & {} \\
             Signing Implementation in Source Code (External Contributors) & Subject's manner of implementing Software signing in first-party source code produced by external contributors like 3rd party vendors, example, commits by contributors are signed   &  S2: "I don't know who these people are ... I got something signed from somebody I don't know ... So what we do though is that we do the code review, and then we have a sign-off by somebody on our team that signs off from that code" \\
             & &\\
             \textbf{DEPLOYMENT/PACKAGE/ARTIFACT} & {} & {} \\
            Signing Implementation in Package/Artifacts&Subject's manner of implementing Software signing in  Deployment of Product or packaging of created artifact (Release) & S10: "git commit signing on the code and then signatures on the final artifact. " \\
            & &\\
             \textbf{THIRD-PARTY DEPENDENCY}&	\\
              Signing Implementation in Third Party Dependencies and Tools & Contains code category related to 1. Influence of Signing in Selection of third party dependencies and 2. Influence of Signing in discontinuation of dependency use	& S13: "So most software package ecosystems that I consume packages from do not support signing consistently enough to put much stock in that. " 
             \\
             & &\\
            \textbf{EFFECT OF SSC FAILURE ON ADOPTION} & {} & {} \\
              Experienced Software Supply Chain Attack & Subject fell victim to a malicious orchestrated software supply chain attack. For example, an organization falling for phishing & S4: "The build system itself was compromised by a state actor...somebody compromised the system and lurked for a long time and built very bespoke malware, specifically tailored to Orion, to mimic Orion's network traffic, all of that other kind of stuff...NSA and GHGQ both said that this was Russian foreign intelligence that did this to SolarWinds." \\
            & &\\
             \textbf{SPECIFIC TOOLING} & {} & {} \\
              Sigstore Criticisms& Criticism and weakness associated with implementing SigStore for signing & S3: "we do have some concerns about the way Fulcio operates as its own certificate authority. So we've been looking at things like OpenPubkey...removing that intermediary...to do identity-based signing directly against the OIDC provider."\\
              & &\\
            \textbf{PERCEIVED IMPORTANCE OF SIGNING} & {} & {} \\
               Provenance and Integrity& Subjects mention that their current or previous teams use software signing in a major role for establishing provenance and integrity or other similar roles. & S4: "A few years from now, everybody expects provenance...they don't install anything if they can't prove where it came from...I think signing is massively important. Signing for open source is probably the single most important thing for software supply chain security" \\
              & &\\
            \textbf{SSC STANDARDS AND REGULATION INFLUENCE} & {} & {} \\
              Influence of Regulations and Standards on SSC Security Strategies & How regulations and standards influence their SSC security strategies on signing, or use of other security methods & S8: "My external looking in perspective of things, is that there's a mix of anticipation of regulation, as well as reaction to regulation. For the most part, there's a lot of trying to predict what is going to come, what the government is seeking for as far as tech regulation, and trying to get ahead of that curve."  \\
             & &\\
             \textbf{CHALLENGES} & {} & {} \\
              Challenges experienced in the use of Software Signing&These are responses from subjects when asked about challenges to them implementing Software Signing generally and not just criticism of a signing implementation like Sigstore/GPG/Notary etc. For example when the subject talks about Key management,  Bureaucracy, motivation & S7: "I'm not confident that those customers actually ever verify those signatures."  \\
            & &\\
             \textbf{OTHERS} & {} & {} \\
             Choice of tool - General Reasons & What influences choice of software supply chain security tools except signing tools & S17: "A lot of the tools are not particularly easy to use "\\
             & &\\
            \bottomrule
		\end{tabular}
		\label{tab:codebook}
	\end{table*}
}
 \pagebreak

\section{Code Evolution Diagram} \label{sec:appendix-CodeEvolutionDiagram}
\begin{figure*}[t]
    \centering
    \includegraphics[width=\linewidth]{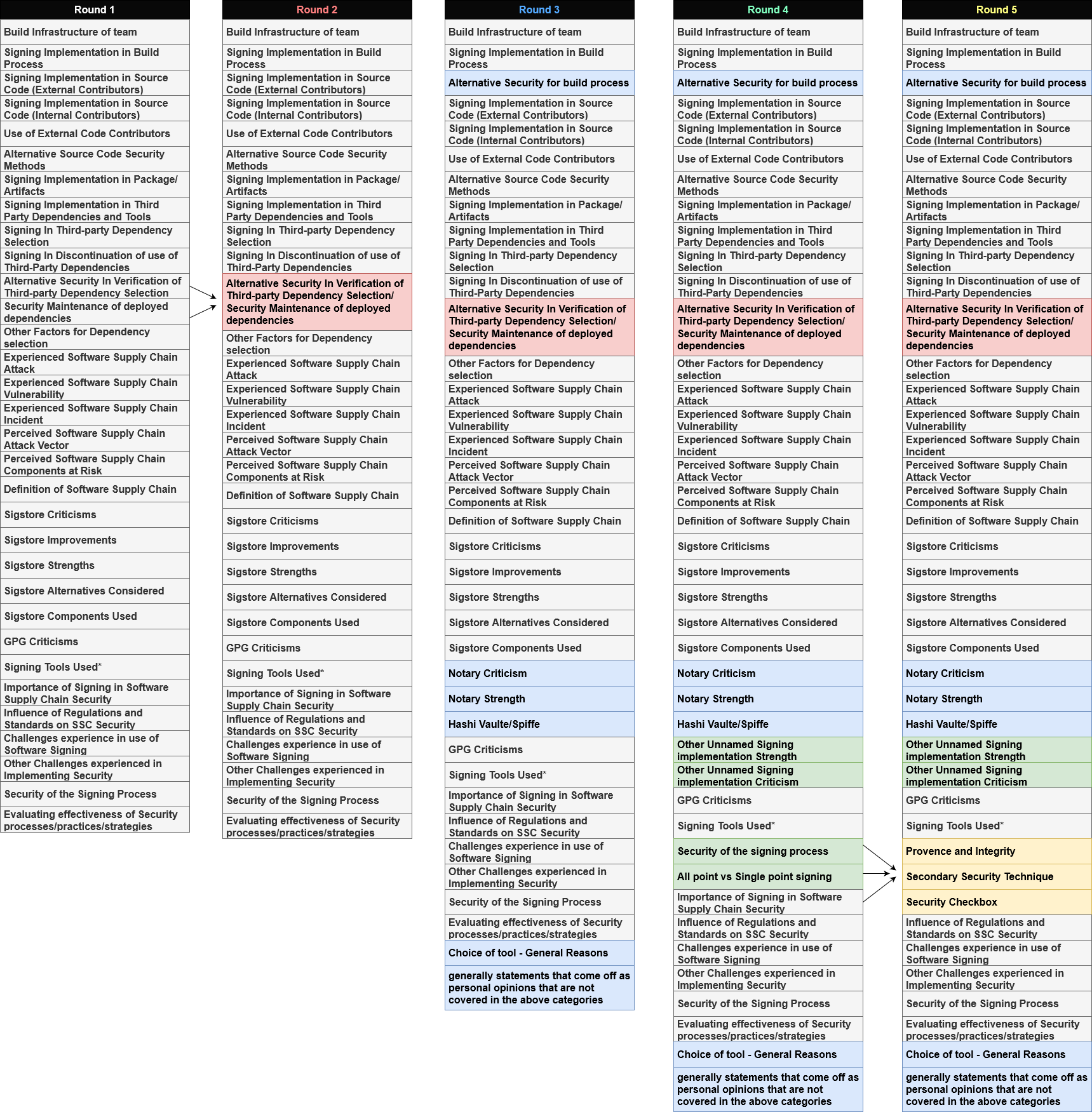}
    \caption{ 
    Codebook Evolution: The codebook underwent five rounds of revisions, with changes highlighted by different colors: round 2 (pink), round 3 (blue), round 4 (green), and round 5 (yellow). Arrows indicate the merger or replacement of codes. The final codes from round 5 were used in this paper, and categorized thematically.
    We combined framework and thematic analyses to develop the codebook. First, we induced codes from each transcript and analyzed them thematically. During the framework analysis, we mapped codes to the four stages/blocks of the software supply chain factory model.
    Round 1: The codebook was defined based on emerging themes.
    Round 2: After the intercoder reliability assessment, two categories were merged.
    Round 3: Blue-highlighted codes (6) were added after memoing and coding the second set of transcripts.
    Round 4: Green-highlighted codes (4) were added.
    Round 5: Analysts A1 and A2 determined that all new codes from stages 3 and 4 (except three(3)) were minor and did not fit the established categories.
    We identified the following themes through framework analysis: Build, Source Code, Deployment, and Third-Party Dependency. Additionally, through thematic analysis, we derived themes such as the Effect of SSC Failures on Signing Adoption, Perceived Importance of Signing, Influence of SSC Standards and Regulations on Signing, and Challenges.
    }
    \label{fig: evolution-fig}
\end{figure*}

~\cref{fig: evolution-fig} describes the evolution of our codebook over five rounds of revisions.

\end{document}

%% file: misc/typesetting.tex
\usepackage{amsmath,amsfonts}
\usepackage{algorithmic}
\usepackage{graphicx}
\usepackage{textcomp}
\usepackage{xcolor}
\usepackage{booktabs} 
\usepackage{xspace} 
\usepackage[normalem]{ulem}
\usepackage{makecell}
\usepackage{tcolorbox}
\usepackage{enumitem}
\usepackage{siunitx}
\usepackage[vskip=1em,font=itshape,leftmargin=2em,rightmargin=2em]{quoting}
\usepackage{svg}
\usepackage{listings}
\usepackage{subfigure}
\usepackage{xcolor}
\PassOptionsToPackage{hyphens}{url}\usepackage{hyperref}

\usepackage{soul}
\usepackage{hyperref}
\usepackage{cleveref}

\ifDEBUG
   \newcommand{\JD}[1]{\textcolor{purple}{[Jamie says: #1]}}
    \newcommand{\KC}[1]{\textcolor{blue}{[Kelechi says: #1]}}
    \newcommand{\TRS}[1]{\textcolor{olive}{[Taylor says: #1]}}
    
    \newcommand{\SA}[1]{\textcolor{orange}{[Santiago says: #1]}}
    \newcommand{\SO}[1]{\textcolor{pink}{[Sofia says: #1]}}
    
    \newcommand{\TODO}[1]{\hl{#1}}
\else
    \newcommand{\JD}[1]{}
    \newcommand{\KC}[1]{}
    \newcommand{\TRS}[1]{}
    \newcommand{\AK}[1]{}
    \newcommand{\SA}[1]{}
    \newcommand{\SO}[1]{}
    \newcommand{\TODO}[1]{#1}
\fi

\newcommand{\myparagraph}[1]{\vspace{0.08cm}\noindent\hspace{0.1cm} \ul{#1:}}

\usepackage{balance} 

\usepackage{amsmath}
\usepackage{cleveref}

\crefformat{section}{\S#2#1#3}
\crefname{figure}{Figure}{Figures}
\crefname{appendix}{Appendix}{Appendices}
\crefname{table}{Table}{Tables}
\crefname{algorithm}{Algorithm}{Algorithms}
\crefname{listing}{Listing}{Listings}
\crefname{theorem}{Theorem}{Theorems}
\crefname{thm}{Theorem}{Theorems}
\crefname{lemma}{Lemma}{Lemmata}
\crefname{equation}{Eqt.}{Eqts.}
\crefformat{Grammar}{Grammar #1}

\usepackage{enumitem}
\usepackage{booktabs}

\usepackage{makecell}
\usepackage{multirow}
\usepackage[T1]{fontenc}

\newcommand{\ie}{\textit{i.e.,} }
\newcommand{\eg}{\textit{e.g.,} }
\newcommand{\etal}{\textit{et al.}\xspace}
\newcommand{\etals}{\textit{et al.'s}\xspace}

\usepackage{enumitem}

\newcommand{\myinlinequote}[1]{\emph{``#1''}}

\usepackage{quoting}

\newcommand{\ParticipantQuote}[2]{
    \begin{quoting}[leftmargin=0.15cm,rightmargin=0.15cm,font=italic]
    \emph{\textbf{#1}: ``#2''} 
    \end{quoting}}

\DeclareUnicodeCharacter{2060}{\nolinebreak}

\newcounter{finding}

\newcommand{\newfinding}{
  \refstepcounter{finding} 
  \textbf{Finding \thefinding.} 
}

%% file: data/data.tex
\newcommand{\maven}{Maven Central\xspace}
\newcommand{\mavenPackages}{{\TODO{30}}\xspace}
\newcommand{\mavenOrganizations}{{\TODO{16}}\xspace}

\newcommand{\mavenReevaluateSelected }{{\TODO{10\%}}\xspace}

\newcommand{\ssc}{Software Supply Chain\xspace}
\newcommand{\sscsm}{software supply chain\xspace}
\newcommand{\sscs}{software supply chains\xspace}
\newcommand{\oss}{Open Source Software\xspace}
\newcommand{\interviews}{18\xspace}
\newcommand{\subjects}{18\xspace}
\newcommand{\sign}{Software signing\xspace}
\newcommand{\signs}{software signing\xspace}
\newcommand{\signature}{Software Signature\xspace}
\newcommand{\orgs}{13\xspace}
\newcommand{\seps}{Software Engineering Processes\xspace}
\newcommand{\sep}{Software Engineering Process\xspace}
\newcommand{\sscfm}{software supply chain factory model\xspace}
\newcommand{\extContribution}{10\xspace}
\newcommand{\reqExternal}{5\xspace}
\newcommand{\intSign}{14\xspace}

\newcommand{\Subjectone}{\emph{S1 \small{\textit{(POC software, digital technology)}}}\xspace}
\newcommand{\Subjecttwo}{\emph{S2 \small{\textit{(SAAS, SSC security tool)}}}\xspace}
\newcommand{\Subjectthree}{\emph{S3 \small{\textit{(SAAS, SSC security tool)}}}\xspace}
\newcommand{\Subjectfour}{\emph{S4 \small{\textit{(open source tooling, cloud dev tool)}}}\xspace}
\newcommand{\Subjectfive}{\emph{S5 \small{\textit{(internal security tool, cloud security)}}}\xspace}
\newcommand{\Subjectsix}{\emph{S6 \small{\textit{(security tooling, digital technology)}}}\xspace}
\newcommand{\Subjectseven}{\emph{S7 \small{\textit{(security tools, cloud security)}}}\xspace}
\newcommand{\Subjecteight}{\emph{S8 \small{\textit{(internal security tool \& cloud APIs, internet services)}}}\xspace}
\newcommand{\Subjectnine}{\emph{S9 \small{\textit{(SAAS, SSC security)}}}\xspace}
\newcommand{\Subjectten}{\emph{S10 \small{\textit{(SAAS, dev tools)}}}\xspace}
\newcommand{\Subjecteleven}{\emph{S11 \small{\textit{(firmware \& testing software, social technology)}}}\xspace}
\newcommand{\Subjecttwelve}{\emph{S12 \small{\textit{(consultancy, cloud OSS security)}}}\xspace}
\newcommand{\Subjectthirteen}{\emph{S13 \small{\textit{(internal security tool, telecommunication)}}}\xspace}
\newcommand{\Subjectfourteen}{\emph{S14 \small{\textit{(POC software \& security tool, cloud security)}}}\xspace}
\newcommand{\Subjectfifteen}{\emph{S15 \small{\textit{(internal security tooling, aerospace security)}}}\xspace}
\newcommand{\Subjectsixteen}{\emph{S16 \small{\textit{(security tooling, digital technology)}}}\xspace}
    \newcommand{\Subjectseventeen}{\emph{S17 \small{\textit{(SAAS, SSC security tool)}}}\xspace}
\newcommand{\Subjecteighteen}{\emph{S18 \small{\textit{(security tooling, internet service)}}}\xspace}

\newcommand{\Subjectoneshort}{\emph{S1}\xspace}
\newcommand{\Subjecttwoshort}{\emph{S2}\xspace}
\newcommand{\Subjectthreeshort}{\emph{S3}\xspace}
\newcommand{\Subjectfourshort}{\emph{S4}\xspace}
\newcommand{\Subjectfiveshort}{\emph{S5}\xspace}
\newcommand{\Subjectsixshort}{\emph{S6}\xspace}
\newcommand{\Subjectsevenshort}{\emph{S7}\xspace}
\newcommand{\Subjecteightshort}{\emph{S8}\xspace}
\newcommand{\Subjectnineshort}{\emph{S9}\xspace}
\newcommand{\Subjecttenshort}{\emph{S10}\xspace}
\newcommand{\Subjectelevenshort}{\emph{S11}\xspace}
\newcommand{\Subjecttwelveshort}{\emph{S12}\xspace}
\newcommand{\Subjectthirteenshort}{\emph{S13}\xspace}
\newcommand{\Subjectfourteenshort}{\emph{S14}\xspace}
\newcommand{\Subjectfifteenshort}{\emph{S15}\xspace}
\newcommand{\Subjectsixteenshort}{\emph{S16}\xspace}
\newcommand{\Subjectseventeenshort}{\emph{S17}\xspace}
\newcommand{\Subjecteighteenshort}{\emph{S18}\xspace}


%% file: main-TechReport-arXiv.bbl
\begin{thebibliography}{10}

\bibitem{noauthor_tag-securitysupply-chain-securitysecure-software-factorysecure-software-factorymd_nodate}
tag-security/supply-chain-security/secure-software-factory/secure-software-factory.md at main · cncf/tag-security.

\bibitem{abbadini_cage4deno_2023}
Marco Abbadini, Dario Facchinetti, Gianluca Oldani, Matthew Rossi, and Stefano Paraboschi.
\newblock {Cage4Deno}: {A} {Fine}-{Grained} {Sandbox} for {Deno} { Subprocesses}.
\newblock In {\em {ACM} AsiaCCS}, 2023.

\bibitem{AJILA20071517}
Samuel~A. Ajila and Di~Wu.
\newblock Empirical study of the effects of open source adoption on software development economics.
\newblock {\em Journal of Systems and Software}, 2007.

\bibitem{alharahsheh2020review}
Husam~Helmi Alharahsheh and Abraham Pius.
\newblock A review of key paradigms: Positivism vs interpretivism.
\newblock {\em Global Academic Journal of Humanities and Social Sciences}, 2020.

\bibitem{Alhindi-2024}
Maysara Alhindi and Joseph Hallett.
\newblock Sandboxing adoption in open source ecosystems.
\newblock In {\em Proceedings of the 12th ACM/IEEE International Workshop on Software Engineering for Systems-of-Systems and Software Ecosystems}, SESoS '24, page 13–20, New York, NY, USA, 2024. Association for Computing Machinery.

\bibitem{amusuo2023ztdjava}
Paschal~C Amusuo, Kyle~A Robinson, Santiago Torres-Arias, Laurent Simon, and James~C Davis.
\newblock Ztdjava: Mitigating software supply chain vulnerabilities via zero-trust dependencies.
\newblock {\em arXiv:2310.14117}, 2023.

\bibitem{anjum_uncovering_2023}
Nafisa Anjum, Nazmus Sakib, Juanjose Rodriguez-Cardenas, Corey Brookins, Ava Norouzinia, Asia Shavers, Miranda Dominguez, Marie Nassif, and Hossain Shahriar.
\newblock Uncovering {Software} {Supply} {Chains} {Vulnerability}: {A} {Review} of {Attack} {Vectors}, {Stakeholders}, and {Regulatory} {Frameworks}.
\newblock In {\em {IEEE} COMPSAC}, 2023.

\bibitem{baltes_sampling_2022}
Sebastian Baltes and Paul Ralph.
\newblock Sampling in software engineering research: a critical review and guidelines.
\newblock {\em Empirical Software Engineering}, 2022.

\bibitem{bogart_when_2021}
Chris Bogart, Christian Kästner, James Herbsleb, and Ferdian Thung.
\newblock When and {How} to {Make} {Breaking} {Changes}: {Policies} and {Practices} in 18 {Open} {Source} {Software} {Ecosystems}.
\newblock {\em ACM Transactions on Software Engineering and Methodology}, 2021.

\bibitem{braun_using_2006}
Virginia Braun and Victoria Clarke.
\newblock Using thematic analysis in psychology.
\newblock {\em Qualitative Research in Psychology}, 2006.

\bibitem{campbell2013coding}
John~L Campbell, Charles Quincy, Jordan Osserman, and Ove~K Pedersen.
\newblock Coding in-depth semistructured interviews: Problems of unitization and intercoder reliability and agreement.
\newblock {\em Sociological methods \& research}, 42(3):294--320, 2013.

\bibitem{cantone_software_1992}
G.~Cantone.
\newblock Software factory: modeling the improvement.
\newblock In {\em 1992 Third International Conference on Factory 2000, 'Competitive Performance Through Advanced Technology'}, pages 124--129, 1992.

\bibitem{chenail2011interviewing}
Ronald~J Chenail.
\newblock Interviewing the investigator: Strategies for addressing instrumentation and researcher bias concerns in qualitative research.
\newblock {\em Qualitative report}, 16(1):255--262, 2011.

\bibitem{cisa_securing_software_supply_chain}
{CISA}.
\newblock Securing the software supply chain for developers.
\newblock Technical report, Cybersecurity and Infrastructure Security Agency, 2022.
\newblock Accessed: 2024-08-28.

\bibitem{cncf_2021}
{Cloud Native Computing Foundation}.
\newblock Software supply chain best practices, May 2021.
\newblock \url{https://github.com/cncf/tag-security/blob/main/supply-chain-security/supply-chain-security-paper/CNCF_SSCP_v1.pdf}.

\bibitem{cncf_supply_chain_security_paper}
{CNCF TAG Security}.
\newblock Cncf software supply chain security paper, 2021.
\newblock Accessed: 2024-08-28.

\bibitem{computer_security_division_secure_2021}
Information Technology~Laboratory Computer Security~Division.
\newblock Secure {Software} {Development} {Framework} {\textbar} {CSRC} {\textbar} {CSRC}, 2021.

\bibitem{cooper_protecting_2018}
David Cooper, Larry Feldman, and Gregory Witte.
\newblock Protecting {Software} {Integrity} {Through} {Code} {Signing}.
\newblock Technical Report ITL Bulletin, National Institute of Standards and Technology, May 2018.

\bibitem{cooper_security_2018}
David Cooper, Andrew Regenscheid, Murugiah Souppaya, et~al.
\newblock Security {Considerations} for {Code} {Signing}.
\newblock {\em NIST Cybersecurity White Paper}, January 2018.

\bibitem{DeJonckheere2019-ow}
Melissa DeJonckheere and Lisa~M Vaughn.
\newblock Semistructured interviewing in primary care research: a balance of relationship and rigour.
\newblock {\em Fam. Med. Community Health}, 2019.

\bibitem{department_of_defense_open_2003}
{Department of Defense}.
\newblock Open {Source} {Software} ({OSS}) in the {Department} of {Defense} ({DoD}), May 2003.

\bibitem{docker_signing_images}
{Docker}.
\newblock Signing docker official images using openpubkey, 2023.
\newblock \url{https://www.docker.com/blog/signing-docker-official-images-using-openpubkey/}.

\bibitem{dodgson2019reflexivity}
Joan~E Dodgson.
\newblock Reflexivity in qualitative research.
\newblock {\em Journal of human lactation}, 2019.

\bibitem{duan_towards_2020}
Ruian Duan, Omar Alrawi, Ranjita~Pai Kasturi, Ryan Elder, Brendan Saltaformaggio, and Wenke Lee.
\newblock Towards {Measuring} {Supply} {Chain} {Attacks} on {Package} {Managers} for {Interpreted} {Languages}, 2020.
\newblock arXiv:2002.01139.

\bibitem{feng_intercoder_2014}
Guangchao~Charles Feng.
\newblock Intercoder reliability indices: disuse, misuse, and abuse.
\newblock {\em Quality \& Quantity}, 2014.

\bibitem{rfc4880}
Hal Finney, Lutz Donnerhacke, Jon Callas, Rodney~L. Thayer, and Daphne Shaw.
\newblock {OpenPGP Message Format}, 2007.

\bibitem{Garfinkel2003-pa}
Simson Garfinkel and Gene Spafford.
\newblock {\em Practical {UNIX} and Internet Security}.
\newblock O'Reilly Media, Sebastopol, CA, 2003.

\bibitem{gokkaya_software_2023}
Betul Gokkaya, Leonardo Aniello, and Basel Halak.
\newblock Software supply chain: review of attacks, risk assessment strategies and security controls, 2023.
\newblock arXiv:2305.14157.

\bibitem{google_software_supply_chain_practices}
{Google Cloud}.
\newblock Best practices for software supply chain security, 2023.
\newblock Accessed: 2024-08-28.

\bibitem{growley_navigating_2021}
Wolff Growley, Lerner Gruden, and Welling Canter.
\newblock Navigating the {SolarWinds} {Supply} {Chain} {Attack}.
\newblock 56(2), 2021.

\bibitem{gsa_18f_18f_nodate}
{GSA: 18F}.
\newblock {18F}: {Digital} service delivery {\textbar} {Open} source policy.
\newblock \url{https://18f.gsa.gov/open-source-policy/}.

\bibitem{guest2006many}
Greg Guest, Arwen Bunce, and Laura Johnson.
\newblock How many interviews are enough? an experiment with data saturation and variability.
\newblock {\em Field methods}, 2006.

\bibitem{Hammi_2023}
Badis Hammi and Sherali Zeadally.
\newblock Software supply-chain security: Issues and countermeasures.
\newblock {\em Computer}, 2023.

\bibitem{heilman_openpubkey_2023}
Ethan Heilman, Lucie Mugnier, Athanasios Filippidis, Sharon Goldberg, Sebastien Lipman, Yuval Marcus, Mike Milano, Sidhartha Premkumar, and Chad Unrein.
\newblock {OpenPubkey}: {Augmenting} {OpenID} {Connect} with {User} held {Signing} {Keys}, 2023.
\newblock Publication info: Preprint.

\bibitem{council2020breaking}
Trey Herr, William Loomis, Stewart Scott, and June Lee.
\newblock Breaking trust: Shades of crisis across an insecure software supply chain, 2020.
\newblock \url{https://www.atlanticcouncil.org/in-depth-research-reports/report/breaking-trust-shades-of-crisis-across-an-insecure-software-supply-chain/}.

\bibitem{holloway1997basic}
Immy Holloway.
\newblock {\em Basic concepts for qualitative research}.
\newblock London ; Malden, MA, USA : Blackwell Science, 1997.
\newblock Includes bibliographical references (pages 164-176).

\bibitem{jiang_empirical_2023}
Wenxin Jiang, Nicholas Synovic, et~al.
\newblock An {Empirical} {Study} of {Pre}-{Trained} {Model} {Reuse} in the {Hugging} {Face} {Deep} {Learning} {Model} {Registry}.
\newblock In {\em {International} {Conference} on {Software} {Engineering} ({ICSE})}, 2023.

\bibitem{kalu_reflecting_2023}
Kelechi Kalu, Taylor~R. Schorlemmer, et~al.
\newblock Reflecting on the {Use} of the {Policy}-{Process}-{Product} {Theory} in {Empirical} {Software} {Engineering}.
\newblock In {\em {ACM} {Joint} {European} {Software} {Engineering} {Conference} and {Symposium} on the {Foundations} of {Software} {Engineering}}, 2023.

\bibitem{kant2023dependencyhell}
Vivek Kant.
\newblock Is software reuse leading to dependency hell?, 2022.
\newblock \url{https://www.linkedin.com/pulse/software-reuse-leading-dependency-hell-vivek-kant/}.

\bibitem{kloeg_charting_2024}
Berend Kloeg, Aaron~Yi Ding, Sjoerd Pellegrom, and Yury Zhauniarovich.
\newblock Charting the {Path} to {SBOM} {Adoption}:{A} {Business} {Stakeholder}-{Centric} {Approach}.
\newblock 2024.

\bibitem{kuppusamy_diplomat_nodate}
Trishank~Karthik Kuppusamy, Santiago Torres-Arias, Vladimir Diaz, and Justin Cappos.
\newblock Diplomat: Using delegations to protect community repositories.
\newblock In {\em USENIX Symposium on Networked Systems Design and Implementation}, 2016.

\bibitem{ladisa_sok_2023}
Piergiorgio Ladisa, Henrik Plate, Matias Martinez, and Olivier Barais.
\newblock {SoK}: {Taxonomy} of {Attacks} on {Open}-{Source} {Software} {Supply} {Chains}.
\newblock In {\em 2023 {IEEE} {Symposium} on {Security} and {Privacy} ({SP})}, 2023.

\bibitem{lamb_reproducible_2022}
Chris Lamb and Stefano Zacchiroli.
\newblock Reproducible {Builds}: {Increasing} the {Integrity} of {Software} {Supply} {Chains}.
\newblock {\em IEEE Software}, 2022.

\bibitem{lamowski_sandcrust_2017}
Benjamin Lamowski, Carsten Weinhold, et~al.
\newblock Sandcrust: {Automatic} {Sandboxing} of {Unsafe} {Components} in { Rust}.
\newblock In {\em {Workshop} on {Prog.} {Langs.} and {Operating} {Syst.} (PLOS)}, 2017.

\bibitem{lauinger_thou_2017}
Tobias Lauinger, Abdelberi Chaabane, et~al.
\newblock Thou {Shalt} {Not} {Depend} on {Me}: {Analysing} the {Use} of {Outdated} {JavaScript} {Libraries} on the {Web}.
\newblock In {\em {Network} and {Distributed} {System} {Security} {Symposium}}, 2017.

\bibitem{li_software_2001}
Chao Li, Huaizhang Li, and Mingshu Li.
\newblock A software factory model based on {ISO9000} and {CMM} for {Chinese} small organizations.
\newblock In {\em Proceedings {Second} {Asia}-{Pacific} {Conference} on {Quality} {Software}}, 2001.

\bibitem{ludvigsen_preventing_2022}
Kaspar~Rosager Ludvigsen, Shishir Nagaraja, and Angela Daly.
\newblock Preventing or {Mitigating} {Adversarial} {Supply} {Chain} {Attacks}: {A} {Legal} {Analysis}.
\newblock In {\em {ACM} {Workshop} on {Software} {Supply} {Chain} {Offensive} {Research} and {Ecosystem} {Defenses}}, 2022.

\bibitem{maxam2024interview}
William~P Maxam~III and James~C Davis.
\newblock An interview study on third-party cyber threat hunting processes in the us department of homeland security.
\newblock {\em USENIX Security}, 2024.

\bibitem{SCIM_Microsoft_2023}
Microsoft.
\newblock microsoft/scim, February 2023.

\bibitem{ntia2021sbom}
{National Telecommunications and Information Administration}.
\newblock The minimum elements for a software bill of materials (sbom), July 2021.
\newblock \url{https://www.ntia.gov/report/2021/minimum-elements-software-bill-materials-sbom}.

\bibitem{neupane_beyond_2023}
Shradha Neupane, Grant Holmes, Elizabeth Wyss, Drew Davidson, and Lorenzo De~Carli.
\newblock Beyond typosquatting: an in-depth look at package confusion.
\newblock In {\em Proceedings of the 32nd {USENIX} {Conference} on {Security} {Symposium}}, {SEC} '23, pages 3439--3456, USA, August 2023. USENIX Association.

\bibitem{newman_sigstore_2022}
Zachary Newman, John~Speed Meyers, and Santiago Torres-Arias.
\newblock Sigstore: {Software} {Signing} for {Everybody}.
\newblock In {\em {ACM} {SIGSAC} {Conference} on {Computer} and {Communications} {Security}}, 2022.

\bibitem{nikiforakis_you_2012}
Nick Nikiforakis, Luca Invernizzi, Alexandros Kapravelos, et~al.
\newblock You are what you include: large-scale evaluation of remote javascript inclusions.
\newblock In {\em {ACM} conference on {Computer} and communications security}, 2012.

\bibitem{oconnor_homesnitch_2019}
TJ~OConnor, Reham Mohamed, Markus Miettinen, William Enck, Bradley Reaves, and Ahmad-Reza Sadeghi.
\newblock {HomeSnitch}: behavior transparency and control for smart home {IoT} devices.
\newblock In {\em Proceedings of the 12th {Conference} on {Security} and {Privacy} in {Wireless} and {Mobile} {Networks}}, {WiSec} '19, pages 128--138, New York, NY, USA, 2019. Association for Computing Machinery.
\newblock event-place: Miami, Florida.

\bibitem{ohm_towards_2020}
Marc Ohm, Arnold Sykosch, and Michael Meier.
\newblock Towards detection of software supply chain attacks by forensic artifacts.
\newblock In {\em ACM {International} {Conference} on {Availability}, {Reliability} and {Security}}, 2020.

\bibitem{okafor_sok}
Chinenye Okafor, Taylor~R. Schorlemmer, Santiago Torres-Arias, and James~C. Davis.
\newblock Sok: Analysis of software supply chain security by establishing secure design properties.
\newblock In {\em ACM Workshop on Software Supply Chain Offensive Research and Ecosystem Defenses}. Association for Computing Machinery, 2022.

\bibitem{openssf_artifact_attestations}
{OpenSSF}.
\newblock Introducing artifact attestations: Now in public beta, 2024.
\newblock \url{https://openssf.org/blog/2024/05/24/introducing-artifact-attestations-now-in-public-beta/}.

\bibitem{oconnor_intercoder_2020}
Cliodhna O’Connor and Helene Joffe.
\newblock Intercoder {Reliability} in {Qualitative} {Research}: {Debates} and {Practical} {Guidelines}.
\newblock {\em International Journal of Qualitative Methods}, 2020.

\bibitem{pashchenko_qualitative_2020}
Ivan Pashchenko, Duc-Ly Vu, and Fabio Massacci.
\newblock A {Qualitative} {Study} of {Dependency} {Management} and {Its} {Security} {Implications}.
\newblock In {\em {ACM} {SIGSAC} {Conference} on {Computer} and {Communications} {Security}}, 2020.

\bibitem{Pigeon_2022}
Daniel Pigeon.
\newblock How to implement 3 new software supply chain security frameworks, Jun 2022.

\bibitem{ruoti2016johnny}
Scott Ruoti, Jeff Andersen, Daniel Zappala, and Kent Seamons.
\newblock Why johnny still, still can't encrypt: Evaluating the usability of a modern pgp client, 2016.
\newblock \url{https://cups.cs.cmu.edu/soups/2006/posters/sheng-poster_abstract.pdf}.

\bibitem{saldana2011fundamentals}
Johnny Saldana.
\newblock {\em Fundamentals of qualitative research}.
\newblock Oxford university press, 2011.

\bibitem{schorlemmer_signing_2024}
T.~R. Schorlemmer, K.~G. Kalu, et~al.
\newblock Signing in four public software package registries: Quantity, quality, and influencing factors.
\newblock In {\em 2024 IEEE Symposium on Security and Privacy (SP)}, 2024.

\bibitem{schorlemmer2025establishing}
Taylor~R Schorlemmer, Ethan~H Burmane, Kelechi~G Kalu, Santiago Torres-Arias, and James~C Davis.
\newblock Establishing provenance before coding: Traditional and next-generation software signing.
\newblock {\em IEEE Security \& Privacy}, 2025.

\bibitem{Internet-Security-Glossary}
R.~Shirey.
\newblock Internet security glossary, version 2.
\newblock Technical report, {RFC} Editor, August 2007.
\newblock \url{https://doi.org/10.17487/rfc4949}.

\bibitem{sigstore_dev}
{Sigstore}.
\newblock Sigstore: A new standard for signing, verifying, and protecting software.
\newblock \url{https://www.sigstore.dev/}.

\bibitem{singi_2019_trusted_ssc}
Kapil Singi, Jagadeesh Chandra~Bose R~P, Sanjay Podder, and Adam~P. Burden.
\newblock Trusted software supply chain.
\newblock In {\em 2019 34th IEEE/ACM International Conference on Automated Software Engineering (ASE)}, 2019.

\bibitem{sommerville_formal_nodate}
Ian Sommerville.
\newblock Formal {Specification}.
\newblock In {\em Software {Engineering}}.

\bibitem{souppaya_secure_2022}
Murugiah Souppaya, Karen Scarfone, and Donna Dodson.
\newblock Secure {Software} {Development} {Framework} ({SSDF}) version 1.1 : recommendations for mitigating the risk of software vulnerabilities.
\newblock Technical report, 2022.

\bibitem{srinivasa_deceptive_2022}
Shreyas Srinivasa, Jens~Myrup Pedersen, and Emmanouil Vasilomanolakis.
\newblock Deceptive directories and “vulnerable” logs: a honeypot study of the {LDAP} and log4j attack landscape.
\newblock In {\em {IEEE} ({EuroS}\&{PW})}, 2022.

\bibitem{srivastava_framework__2009}
Aashish Srivastava and S.~Bruce Thomson.
\newblock Framework {Analysis}: {A} {Qualitative} {Methodology} for {Applied} {Policy} {Research}, January 2009.
\newblock \url{https://papers.ssrn.com/abstract=2760705}.

\bibitem{stafford2020zero}
V~Stafford.
\newblock Zero trust architecture.
\newblock {\em NIST Special Publication}, 800:207, 2020.
\newblock \url{https://nvlpubs.nist.gov/nistpubs/SpecialPublications/NIST.SP.800-207.pdf}.

\bibitem{sun_nativeguard_2014}
Mengtao Sun and Gang Tan.
\newblock {NativeGuard}: protecting android applications from third-party native libraries.
\newblock In {\em {ACM} WiSec}, 2014.

\bibitem{ossra-2020}
Synopsys.
\newblock {2020 Open Source Security and Risk Analysis (OSSRA) Report}, 2020.
\newblock \url{ https://www.synopsys.com/software-integrity/resources/analyst-reports/2020-open-source-security-risk-analysis.html }.

\bibitem{OSSRA_2024}
Synopsys.
\newblock {2024 Open Source Security and Risk Analysis (OSSRA) Report}, 2024.
\newblock \url{ https://www.synopsys.com/software-integrity/resources/analyst-reports/open-source-security-risk-analysis.html#introMenu }.

\bibitem{terry2017thematic}
Gareth Terry, Nikki Hayfield, Victoria Clarke, Virginia Braun, et~al.
\newblock Thematic analysis.
\newblock {\em The SAGE handbook of qualitative research in psychology}, 2(17-37):25, 2017.

\bibitem{Thabane_Ma_Chu_Cheng_Ismaila_Rios_Robson_Thabane_Giangregorio_Goldsmith_2010}
Lehana Thabane, Jinhui Ma, Rong Chu, Ji~Cheng, Afisi Ismaila, Lorena~P. Rios, Reid Robson, Marroon Thabane, Lora Giangregorio, and Charles~H. Goldsmith.
\newblock A tutorial on pilot studies: the what, why and how.
\newblock {\em BMC Medical Research Methodology}, 10(1):1, January 2010.

\bibitem{SLSA_The_Linux_Foundation_2023}
{The Linux Foundation}.
\newblock Supply-chain levels for software artifacts.
\newblock \url{https://slsa.dev/}.

\bibitem{whitehouse_cybersecurity_2021}
{The White House}.
\newblock Executive order on improving the nation’s cybersecurity, May 2021.
\newblock \url{https://www.whitehouse.gov/briefing-room/statements-releases/2021/05/12/executive-order-on-improving-the-nations-cybersecurity/}.

\bibitem{intoto}
Santiago Torres-Arias, Hammad Afzali, Trishank~Karthik Kuppusamy, Reza Curtmola, and Justin Cappos.
\newblock in-toto: Providing farm-to-table guarantees for bits and bytes.
\newblock In {\em 28th USENIX Security Symposium (USENIX Security 19)}, pages 1393--1410, Santa Clara, CA, August 2019. USENIX Association.

\bibitem{vasilakis_supply-chain_2021}
Nikos Vasilakis, Achilles Benetopoulos, Shivam Handa, Alizee Schoen, Jiasi Shen, and Martin~C. Rinard.
\newblock Supply-{Chain} {Vulnerability} {Elimination} via {Active} {Learning} and {Regeneration}.
\newblock In {\em Proceedings of the 2021 {ACM} {SIGSAC} {Conference} on {Computer} and {Communications} {Security}}, {CCS} '21, pages 1755--1770, New York, NY, USA, November 2021. Association for Computing Machinery.

\bibitem{vasilakis_breakapp_2018}
Nikos Vasilakis, Ben Karel, Nick Roessler, Nathan Dautenhahn, Andre DeHon, and Jonathan~M. Smith.
\newblock {BreakApp}: {Automated}, {Flexible} {Application} { Compartmentalization}.
\newblock In {\em NDSS}, 2018.

\bibitem{verdecchia2023threats}
Roberto Verdecchia, Emelie Engstr{\"o}m, Patricia Lago, Per Runeson, and Qunying Song.
\newblock Threats to validity in software engineering research: A critical reflection.
\newblock {\em Information and Software Technology}, 164:107329, 2023.

\bibitem{voils2007or}
Corrine~I Voils, Julie Barroso, Victor Hasselblad, and Margarete Sandelowski.
\newblock In or out? methodological considerations for including and excluding findings from a meta-analysis of predictors of antiretroviral adherence in hiv-positive women.
\newblock {\em Journal of Advanced Nursing}, 59(2):163--177, 2007.

\bibitem{vu_lastpymile_2021}
Duc-Ly Vu, Fabio Massacci, Ivan Pashchenko, Henrik Plate, and Antonino Sabetta.
\newblock {LastPyMile}: identifying the discrepancy between sources and packages.
\newblock In {\em {ACM} {Joint} {Meeting} on {European} {Software} {Engineering} {Conference} and {Symposium} on the {Foundations} of {Software} {Engineering}}, Athens Greece, 2021. ACM.

\bibitem{wan2019practical}
Zhiyuan Wan, David Lo, Xin Xia, and Liang Cai.
\newblock Practical and effective sandboxing for linux containers.
\newblock {\em Empirical Software Engineering}, 24:4034--4070, 2019.

\bibitem{wermke_committed_2022}
Dominik Wermke, Noah Wohler, Jan~H. Klemmer, Marcel Fourne, Yasemin Acar, and Sascha Fahl.
\newblock Committed to {Trust}: {A} {Qualitative} {Study} on {Security} \& {Trust} in {Open} {Source} {Software} {Projects}.
\newblock In {\em {IEEE} {Symposium} on {Security} and {Privacy}}, 2022.

\bibitem{wheeler_2023_SLSA_survey}
David~A. Wheeler, John~Speed Meyers, Mikael Barbero, and Rebecca Rumbul.
\newblock New slsa++ survey reveals real-world developer approaches to software supply chain security, 2023.
\newblock \url{ https://openssf.org/blog/2023/03/15/new-slsa-survey-reveals-real-world-developer-approaches-to-software-supply-chain-security/ }.

\bibitem{whitten1999johnny}
Alma Whitten and J~Doug Tygar.
\newblock Why johnny can't encrypt: A usability evaluation of pgp 5.0.
\newblock In {\em USENIX security symposium}, volume 348, pages 169--184, 1999.

\bibitem{wittern2016look}
Erik Wittern, Philippe Suter, and Shriram Rajagopalan.
\newblock A look at the dynamics of the javascript package ecosystem.
\newblock In {\em Proceedings of the 13th international conference on mining software repositories}, pages 351--361, 2016.

\bibitem{wohlin2012experimentation}
Claes Wohlin, Per Runeson, Martin H{\"o}st, Magnus~C Ohlsson, Bj{\"o}rn Regnell, Anders Wessl{\'e}n, et~al.
\newblock {\em Experimentation in software engineering}, volume 236.
\newblock Springer, 2012.

\bibitem{woodruff_pgp_2023}
William Woodruff.
\newblock {PGP} signatures on {PyPI}: worse than useless, May 2023.
\newblock \url{ https://blog.yossarian.net/2023/05/21/PGP-signatures-on-PyPI-worse-than-useless }.

\bibitem{xia_empirical_2023}
Boming Xia, Tingting Bi, Zhenchang Xing, Qinghua Lu, and Liming Zhu.
\newblock An {Empirical} {Study} on {Software} {Bill} of {Materials}: {Where} {We} {Stand} and the {Road} {Ahead}.
\newblock In {\em {IEEE}/{ACM} {International} {Conference} on {Software} {Engineering}}, 2023.

\bibitem{xia_trust_2023}
Boming Xia, Dawen Zhang, Yue Liu, Qinghua Lu, Zhenchang Xing, and Liming Zhu.
\newblock Trust in {Software} {Supply} {Chains}: {Blockchain}-{Enabled} {SBOM} and the {AIBOM} {Future}, 2023.
\newblock arXiv:2307.02088.

\bibitem{yan_estimating_2021}
Dapeng Yan, Yuqing Niu, Kui Liu, Zhe Liu, Zhiming Liu, and Tegawende~F. Bissyande.
\newblock Estimating the {Attack} {Surface} from {Residual} {Vulnerabilities} in {Open} {Source} {Software} {Supply} {Chain}.
\newblock In {\em {International} {Conference} on {Software} {Quality}, {Reliability} and {Security} ({QRS})}. IEEE, 2021.

\bibitem{zahan2023openssf}
Nusrat Zahan, Parth Kanakiya, Brian Hambleton, Shohanuzzaman Shohan, and Laurie Williams.
\newblock Openssf scorecard: On the path toward ecosystem-wide automated security metrics.
\newblock {\em IEEE Security \& Privacy}, 21(6):76--88, 2023.

\bibitem{zahan_what_2022}
Nusrat Zahan, Thomas Zimmermann, Patrice Godefroid, Brendan Murphy, Chandra Maddila, and Laurie Williams.
\newblock What are {Weak} {Links} in the npm {Supply} {Chain}?
\newblock In {\em {IEEE}/{ACM} ICSE-SEIP}, 2022.

\bibitem{zimmermann_small_nodate}
Markus Zimmermann, Cristian-Alexandru Staicu, Cam Tenny, and Michael Pradel.
\newblock Small world with high risks: A study of security threats in the npm ecosystem.
\newblock In {\em USENIX Security Symposium}. USENIX Association, 2019.

\bibitem{miranda_zottmann_comparing_2023}
Miranda Zottmann, Carlos Eduardo, Stuckert do~Amaral, Thiago Melo, et~al.
\newblock Comparing {Software} {Supply} {Chain} {Protection} {Approaches}.
\newblock In {\em {Workshop} on {Communication} {Networks} and {Power} {Systems} ({WCNPS})}, 2023.

\end{thebibliography}
